\documentclass[utf8]{FrontiersinHarvard} 

\usepackage{url,hyperref,lineno,microtype,subcaption}
\usepackage[onehalfspacing]{setspace}

\def    \apjl  		{\rm {ApJL}}
\def    \apj  		{\rm {ApJ}}
\def    \apjs                {\rm {ApJS}}
\def    \aj                    {\rm {AJ}}
\def    \mnras  	        {\rm {MNRAS}}
\def    \araa  		{\rm {ARA\&A}}
\def    \aap  		{\rm {A\&A}}
\def    \apjl  		{\rm {ApJL}}

\def    \apss              {\rm {AP$\&$SS}}
\def    \pasj               {\rm {PASJ}}
\def    \jqsrt               {\rm {JQSRT}}
\def    \ao                  {\rm {ApOpt}}
\def    \qjras                  {\rm {QJRAS}}
\def    \pasp                  {\rm {PASP}}

\def\keyFont{\fontsize{8}{11}\helveticabold }
\def\firstAuthorLast{Tram \& Hoang} 
\def\Authors{Le Ngoc Tram\,$^{1}$ and Thiem Hoang\,$^{2,3}$}
\def\Address{$^{1}$Max-Planck-Institut f\"ur Radioastronomie, Auf dem H\"ugel 69, 53121, Bonn, Germany \\
$^{2}$Korea Astronomy and Space Science Institute, Daejeon 34055, South Korea \\ 
$^{3}$Korea University of Science and Technology, 217 Gajeong-ro, Yuseong-gu, Daejeon, 34113, South Korea}
\def\corrAuthor{Corresponding Authors: Le Ngoc Tram and Thiem Hoang}

\def\corrEmail{nle@mpifr-bonn.mpg.de and thiemhoang@kasi.re.kr}

\begin{document}
\onecolumn
\firstpage{1}

\title[Recent progress in theory and observations of grain alignment and disruption]{Recent progress in theory and observational study of dust grain alignment and rotational disruption in star-forming regions} 

\author[\firstAuthorLast ]{\Authors} 
\address{} 
\correspondance{} 

\extraAuth{}

\maketitle

\begin{abstract}
Modern understanding of dust astrophysics reveals that RAdiative Torques (RATs) arising from the radiation-dust interaction can induce two fundamental effects, including grain alignment and rotational disruption. Here we review the recent progress in the theoretical development and observational testing of these effects using dust polarization observed toward star-forming regions (SFRs). We first review the basic theory of the RAT alignment and RAT disruption, which are referred to as RAT-A and RAT-D effects, respectively. We then briefly describe the numerical method we use to model polarized thermal dust emission by accounting for both RAT-A and RAT-D and theoretical predictions of dust polarization for observations. Next, we review our observational efforts to search for observational evidence of the RAT-A and RAT-D effects using thermal dust polarization toward SFRs. Finally, we discuss magnetic fields inferred from dust polarization observed toward these SFRs and implications of the RAT paradigm for different astrophysical conditions, including protostellar environments, dust evolution, and time-domain astrophysics. 
\section{}

\tiny
 \keyFont{ \section{Keywords:} Interstellar medium; Astrophysical dust processes; Interstellar dust; Interstellar dust extinction; Dust continuum emission; Starlight polarization; Interstellar magnetic fields; Star formation.}
\end{abstract}

\section{Introduction}
Dust is an essential component of the interstellar medium (ISM) and plays an important role in many astrophysical phenomena, from the ISM evolution, star formation, planet formation to surface chemistry. The dust properties, including the size, shape, and composition, determine the extinction of starlight and reemission in infrared emission, which controls the appearance of the Universe (\citealt{2011piim.book.....D}). Thermal dust emission is at the foundation of infrared, submm/mm astronomy, a central topic of which is the formation process of stars and planets. In particular, the discovery of starlight polarization (\citealt{Hall:1949p5890,Hiltner:1949p5856}) revealed that dust grains are non-spherical and aligned with a preferred direction in space, which demonstrated the existence of the interstellar magnetic field.

Magnetic field is believed to play an important role in various astrophysical phenomena, including the evolution of the ISM and star formation \citep{2012ARA&A..50...29C}, cosmic ray acceleration and transport. Therefore, observations of magnetic field have important implications for understanding astrophysical phenomena. Motivated by the discovery of starlight polarization and grain alignment, \cite{1951PhRv...81..890D} and \cite{1953ApJ...118..113C} introduced a technique to measure the field strength using dust polarization orientations, which is based on the energy balance between turbulent energy and magnetic energy (known as DCF technique). \cite{Hildebrand:1988p2566} realized that polarization of far-IR thermal emission from dust grains aligned with the magnetic field can be used to trace the field morphology in very cold regions where young stars are being formed because long-wavelength photons can propagate a long distance. These enabled the rapid development of far-IR/submm polarimetric capabilities in the last decades, including SMA (\citealt{2008SPIE.7020E..2BM}), CSO/SHARP (\citealt{2008ApOpt..47..422L}), {\it Planck} (\citealt{2014A&A...571A...1P}), BLAST-Pol (\citealt{2014SPIE.9145E..0RG}); PILOT (\citealt{2014SPIE.9153E..1HM}), APEX/POLKA (\citealt{2014PASP..126.1027W}), CARMA (\citealt{2015JAI.....450005H}), ALMA (\citealt{2016ApJ...824..132N}), JCMT/POL2 (\citealt{2018SPIE10708E..3MF}), and SOFIA/HAWC+ (\citealt{2018JAI.....740008H}). As a result, significant progress in understanding the role of magnetic fields in the interstellar process have been made, from the diffuse interstellar clouds (e.g., \citealt{2005LNP...664..137H};\citealt{refId0}), star-forming regions (e.g., \citealt{2007MNRAS.382..699C,2019ApJ...872..187C,2020ApJ...899...28D,2021A&A...647A..78A,2021ApJ...911...81D,2021ApJ...908...98G,2022ApJ...926..163K}, see also \citealt{2022arXiv220311179P} for reviews), to extra-galactic scales (e.g., \citealt{2020ApJ...888...66L,2021NatAs...5..604L};\citealt{2020AJ....160..167J}; \citealt{2021ApJ...921..128B}; \citealt{2021MNRAS.505..684P}).

However, the question of how dust grains get aligned and whether dust polarization is a reliable tracer of magnetic field is a long-standing question in astrophysics (see \citealt{2007JQSRT.106..225L} for a review). In principle, the process of grain alignment includes two stages, (1) the alignment of the grain axis of maximum inertia moment (i.e., the shortest axis) with its angular momentum (the so-called internal alignment), and (2) the alignment of the grain angular momentum with a preferred axis in the space (i.e., axis of grain alignment), which includes the external magnetic field and radiation direction (the so-called external alignment). 
The leading process for the internal alignment of interstellar grains is the Barnett relaxation effect (\citealt{barnett_s_j_1909_1448026,1979ApJ...231..404P}). The Barnett effect is the reverse of the Einstein-de-Haas effect (\citealt{1915KNAB...18..696E}), arising from the dissipation of grain rotational energy into heat due to rotating magnetization within a paramagnetic grain rotating around a non-principal axis. 

For the ISM (e.g., the diffuse ISM and molecular clouds (MCs)) where grains are essentially small (of radius of $a<1\, \rm \mu m$), the Barnett relaxation is usually much faster than the randomization of grain orientation by gas collisions, resulting in the perfect internal grain alignment (i.e., the grain axis of maximum inertia moment is parallel to its angular momentum; see e.g., \citealt{2022ApJ...928..102H}). For external alignment, the leading theory of grain alignment is based on RAdiative Torques (hereafter RATs, \citealt{1976Ap&SS..43..291D}; \citealt{1996ApJ...470..551D}; \citealt{2007MNRAS.378..910L}) and/or mechanical torques (METs; \citealt{2007ApJ...669L..77L,2018ApJ...852..129H}. Many predictions of the RAT alignment (or RAT-A) theory were successfully tested with observations \citep{2015ARA&A..53..501A}. The fast Larmor precession establishes the magnetic field as a preferred axis of the external alignment (see e.g., \citealt{2022ApJ...928..102H}). Therefore, dust polarization is an established, reliable tracer of magnetic fields in such conditions, which is the focus of this review.

Yet, for special environments where the radiation intensity is strong, while magnetic field is weak or grain susceptibility is negligible (e.g., carbonaceous dust), or the gas is very dense (e.g., protostellar cores/disks and protoplanetary disks, in which grains could grow to very large size of $>10\,\mu$m), the physics of grain alignment becomes more complex. For example, large grains may be aligned with the radiation direction instead of magnetic field, as originally suggested in \cite{2007MNRAS.378..910L} and modeled in \cite{2017ApJ...839...56T} and \cite{2021MNRAS.503.3414P}. As a result, it is not clear whether dust polarization can still trace magnetic fields in such very dense environments. A detailed study of grain alignment for large grains in protostellar environments is presented in \cite{Hoang.2022}.

While dust polarization orientations become a powerful method to measure magnetic field, the dust polarization degree ($p$) provides a unique probe into dust properties and dust fundamental physics. Indeed, the degree of dust polarization depends on the grain alignment efficiency, grain properties, the local density and magnetic field structure (e.g., \citealt{2009ApJ...696....1D}; \citealt{2018A&A...610A..16G}). According to the RAT-A theory, the alignment efficiency of dust grains depends on the local conditions (gas properties and radiation, \citealt{2021ApJ...908..218H}), grain geometric properties (size and shape) and magnetic properties \citep{2007MNRAS.378..910L,2008MNRAS.388..117H,2016ApJ...831..159H,2021ApJ...913...63H}. Therefore, $p$ allows to probe physics of grain alignment and physical properties of dust grains such as grain sizes and composition. The variation of $p$ with the total emission intensity ($I$) permits a popular analysis of polarimetric data to study the variation of grain alignment and the magnetic field geometry across the observed region. The general trend observed toward molecular clouds is $p\propto I^{-\alpha}$ with the power-index $\alpha$ between $0$ and $1$ (e.g., \citealt{2019ApJ...877...88C}; \citealt{2019ApJ...880...27P}; \citealt{2021ApJ...908...10N}). If the magnetic field is uniform, $\alpha=0$ indicating that grain alignment is uniform, while $\alpha=1$ implying that grain alignment only occurs in the outer layer of the cloud and becomes completely lost in the inner region (\citealt{2008ApJ...674..304W}). If magnetic fields are completely random, the uniform grain alignment results in a slope of $\alpha \approx 0.5$ (see \citealt{1992ApJ...389..602J,2015AJ....149...31J}). 

Nevertheless, $p-I$ relation is not a perfect tracer of grain alignment because of the complex dependence of the emission intensity on the gas density and dust properties. Indeed, the total emission intensity observed at frequency $\nu$ is $I_{\nu} \propto B(T_{\rm d})\tau_{\nu} \sim f_{d/g}N_{\rm H}\times B(T_{\rm d})\times \kappa_{\nu} $ with $f_{d/g}$ the dust-to-gas mass ratio and $\kappa_{\nu}$ the dust opacity, which is the product of the gas column density ($N_{\rm H}$), dust temperature ($T_{\rm d}$), and dust opacity (i.e., grain size distribution and composition) along the line of sight. Therefore, the $p-I$ relation reflects the overall dependence of $p$ on the product of three key parameters (gas density, radiation field, and grain size distribution). To disentangle the effect of the radiation field from the gas density, it is required to analyze the variation of the polarization degree with the dust temperature and the gas column density. The first analysis directly reveals the grain alignment 
by RATs, while the second one reflects the effect of grain randomization induced by gas
collisions. The dust polarization is predicted to increase with radiation emission intensity (or $T_{\rm d}$) by theory (\citealt{2021ApJ...908..218H}), mumerical modeling (\citealt{2020ApJ...896...44L}) and numerical simulations (\citealt{2016A&A...593A..87R}). A detailed analysis of the polarization data for various clouds by \cite{2020A&A...641A..12P} shows that the polarization degree at 850 $\mu$m decreases with dust temperature for $T_{\rm d}\geq 19\,\rm K$ (see their Figure 18). This observed feature challenges the classical picture of the RAT-A theory.

Recently, a new fundamental effect induced by RATs, so-called Radiative Torque Disruption (or RAT-D), has been realized by \cite{2019NatAs...3..766H,2019ApJ...876...13H}. The effect is based on the fact that an irregular dust grain exposed to a strong radiation field could be disrupted into small fragments when the centrifugal stress induced by suprathermal
rotation by RATs exceeds the maximum binding energy of the grain material (i.e., tensile strength). Since RATs are stronger for larger grains (\citealt{2007MNRAS.378..910L}; \citealt{2008MNRAS.388..117H}), RAT-D is effective for the large grains, which then determines the upper cutoff of the grain size distribution (GSD) and causes the variation of GSD with the radiation field. The modified GSD by RAT-D in turn, affects dust optical-UV extinction, thermal emission from infrared to radio wavelengths, as well as polarization (see \citealt{2020Galax...8...52H} for a review). The RAT-D effect has been studied in many astrophysical environments, including extinction curves toward starburst galaxies (\citealt{2020MNRAS.494.1058H}), light-curve of type Ia supernovae (\citealt{2020ApJ...888...93G}) and of Gamma-Ray Burst (GRB) Afterglows (\citealt{2020ApJ...895...16H}), evolution of ice in protoplanetary disks (\citealt{2020ApJ...901....6T}) and cometary comae (\citealt{2020ApJ...901...59H}), circumsolar dust (\citealt{2021ApJ...919...91H}), rotational desorption of interstellar organic molecules (\citealt{2020ApJ...891...38H}), spinning dust nanoparticles in circumstellar envelopes of evolved stars (\citealt{2020ApJ...893..138T}), radiation pressure feedback from massive protostars \citep{Hoang.2021a} and dust evolution across cosmic time \citep{Hoang.2021b}. In particular, \cite{2020ApJ...896...44L} performed numerical modeling of multi-wavelength polarized thermal dust emission from aligned grains and disruption by RATs in MCs. They showed that the polarization at 850$\,\mu$m first increases and then decreases with $T_{\rm d}$, which could reproduce the $p-T_{\rm d}$ trend observed by {\it Planck} (\citealt{2020A&A...641A..12P}). \cite{2021ApJ...906..115T} carried out a detailed modelling to interpret the $p-T_{\rm d}$ of Ophiuchus-A cloud observed by SOFIA/HAWC+ with higher resolution than {\it Planck}. The authors demonstrated that the combination of the RAT-A and RAT-D effects could successfully reproduce the observational data. Since then, more observations have shown evidences for the RAT-D mechanism (e.g., \citealt{2021ApJ...908...10N}; \citealt{2021ApJ...923..130T}; \citealt{2022ApJ...929...27H}).

High-resolution polarimetric observations toward starless cores (e.g.,\citealt{2014A&A...569L...1A}; \citealt{2015AJ....149...31J}) and protostars (e.g., \citealt{2014ApJS..213...13H}; \citealt{2018ApJ...855...92C}; \citealt{2019FrASS...6...15P}) usually report the decrease of the polarization degree with increasing column density or intensity of dust emission, a.k.a “polarization hole”. Two popular reasons proposed to explain the polarization hole includes the decrease of grain alignment efficiency toward denser gas/lower radiation field and the tangling of the magnetic field (\citealt{1992ApJ...399..108G,1995ApJ...448..748G}; \citealt{2014ApJS..213...13H}). In the case of protostars, one expects, however, an increase of grain alignment efficiency toward the protostar due to the increasing incident radiation flux with accordance to the RAT-A theory. Hence, an increase in the induced polarization degree is expected in proximity to the peak intensity. Therefore, the underlying origin of the polarization hole toward protostars is difficult to understand and reconcile in terms of grain alignment theory. The miss-match between theory and observations has been reported in \cite{2020NatAs...4.1195P} and \cite{2021A&A...645A.125Z}. \cite{2021ApJ...908..218H} showed that the rotational disruption by RATs could occur in the vicinity of protostars and suggested the combination of grain alignment and disruption by RATs as a plausible origin of the polarization hole. For MCs without embedded sources or line of sights far away from the source, the polarization hole may be dominated by the effect of the field tangling and grain alignment loss (see \citealt{2019FrASS...6...15P}).

In this review, we summarize the fundamentals of grain alignment in the context of suprathermal rotation and disruption by RATs, internal and external alignment processes within conditions of molecular cloud in Section \ref{sec:RAT_paradigm}. In Section \ref{sec:modelling}, we summarize the numerical modelling that predicts the polarized thermal dust emission of molecular cloud irradiated by a nearby luminosity source in the framework of the RAT paradigm (RAT-A and RAT-D effects). We review some representative observational interpretations in Section \ref{sec:obs}. In Section \ref{sec:discussion}, we discuss grain alignment to trace magnetic fields in molecular clouds and the very dense conditions in protostellar cores/disks, including polarization hole within the RAT paradigm, effects of RAT paradigm in time-domain astrophysics and on grain growth and evolution, and suprathermal rotation induced by mechanical torques.

\section{Grain Alignment and Disruption by Radiative Torques: the RAT paradigm}\label{sec:RAT_paradigm}

\subsection{Radiative torques and its fundamental effects}
\cite{1976Ap&SS..43..291D} first suggested that the interaction of an anisotropic radiation with a helical grain can induce RATs due to the differential scattering/absorption of left- and right-handed circularly polarized photons. Latter, RATs were numerically demonstrated in \cite{1996ApJ...470..551D} for three irregular grain shapes. \cite{2007MNRAS.378..910L} introduced an Analytical Model (AMO) of RATs which is based on a helical grain consisting of an oblate spheroid and a weightless mirror. The AMO is shown to reproduce the basic properties of RATs obtained from numerical calculations for realistically irregular grain shapes \citep{2007MNRAS.378..910L,2008MNRAS.388..117H,2021ApJ...913...63H}, and enables us to make quantitative predictions for various conditions \citep{2014MNRAS.438..680H} and dust compositions \citep{2008ApJ...676L..25L,2009ApJ...695.1457H,2016ApJ...831..159H}. Many predictions from the RAT alignment (hereafter RAT-A) theory were observationally tested (see \citealt{2015ARA&A..53..501A} \citealt{2015psps.book...81L} for recent reviews).

As shown in previous studies \citep{1997ApJ...480..633D,2007MNRAS.378..910L,2008MNRAS.388..117H}, RATs in general can induce three fundamental effects on grain rotational dynamics, including (1) the grain precession around the radiation direction, (2) spin-up the grain to suprathermal rotation as well as spin-down to thermal rotation, and (3) align the grain with $\boldsymbol{J}$ along the radiation $\boldsymbol{k}$.

\subsection{Grain model and magnetic properties}
\subsubsection{Grain model}
\begin{figure}
    \centering
    \includegraphics[width=0.9\textwidth]{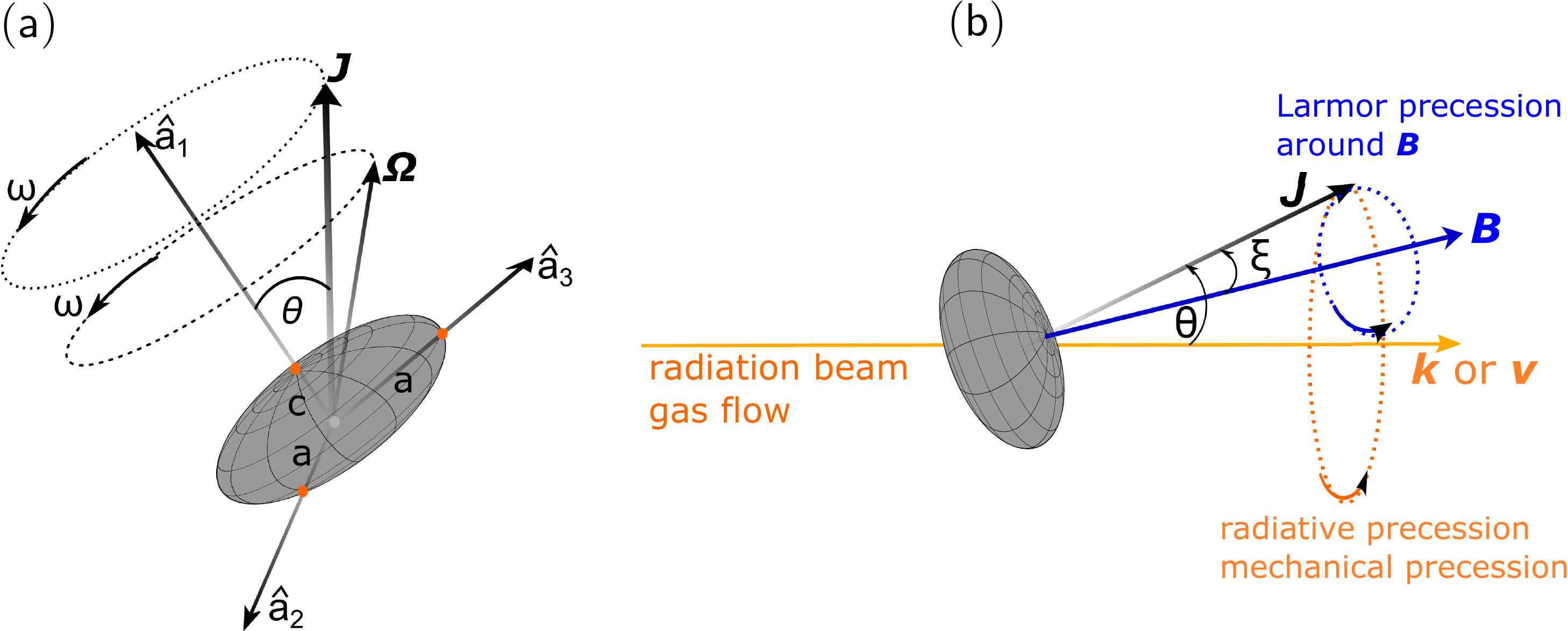}
    \caption{\textbf{(a)}: illustration of internal alignment between the angular momentum ($\boldsymbol{J}$) and the short axis ($\boldsymbol{a_{1}}$). \textbf{(b)}: illustration of external alignment between $\boldsymbol{J}$ and the external magnetic field ($\boldsymbol{B}$) or radiation direction ($\boldsymbol{k}$) or gas flow ($\boldsymbol{v}$). $\boldsymbol{\Omega}$ is the angular velocity, and $\omega$ is the angular frequency. Figures taken from \cite{Hoang.2022}.}
    \label{fig:grain_define}
\end{figure}

Let us define an irregular grain of size $a$ and mass density $\rho$ (Figure \ref{fig:grain_define}). In the fixed grain's frame, this irregular shape is described by there principal axes $\hat{a}_{1},\hat{a}_{2},\hat{a}_{3}$ with the inertia moments of $I_{1}> I_{2}\ge I_{3}$, respectively. For the sake of simplicity, we assume an oblate spheroidal shape (i.e., $I_{1}> I_{2}=I_{3}$) with the length of the semi-minor axis of $c$ and the lengths of the semi-major axes of $a$. The corresponding inertial moments are $I_{\|}\equiv I_{1}=\frac{8\pi}{15}\rho a^{4}c=\frac{8\pi}{15}\rho sa^{5}$ and $I_{\perp}\equiv I_{2}=I_{3}=\frac{4\pi}{15}\rho a^{2}c\left(a^{2}+c^{2}\right)=\frac{4\pi}{15}\rho sa^{5}\left(1+s^{2}\right)$, where $s=c/a$ is the axial ratio of oblate grains. The volume of this grain is $V=4/3\pi sa^{3}$. 

For an isolated grain, the angular momentum $\boldsymbol{J}$ is conserved, and the grain axis of maximum inertia in general makes an angle with $\boldsymbol{J}$. Figure \ref{fig:grain_define}(a) describes the torque-free motion of the oblate grain, which includes the precession of angular velocity of $\boldsymbol{\Omega}$ or momentum $\boldsymbol{J}$ with a frequency of $\omega$ along the grain shortest axis. In the presence of an external magnetic field, radiation field or a gas flow, the angular momentum $\boldsymbol{J}$
precesses around the magnetic field, radiation field or incident gas flow, Figure \ref{fig:grain_define}(b) \citep{2007MNRAS.378..910L,2007ApJ...669L..77L}.

\subsubsection{Grain magnetic properties}\label{sec:susceptibility}
Magnetic properties of dust are essential for their interaction with the ambient magnetic field and grain alignment. We consider grains having embedded iron atoms (such as silicate grains). The presence of iron atoms with unpaired electrons makes interstellar dust a natural paramagnetic material. When iron atoms are diffusely distributed, the grain is called paramagnetic (PM). When iron atoms are distributed as iron clusters, dust grains become superparamagnetic material (SPM, see \citealt{2016ApJ...831..159H}).

The magnetic susceptibility of a paramagnetic grain at rest is given by
\begin{equation}
    \chi_{\rm PM}(0) = 0.03n_{23}f_{p}\hat{p}^{2}\left(\frac{20\,\rm K}{T_{\rm d}}\right)~{,}
\end{equation}
with $\hat{p}=p/5.5$ ($p\simeq 5.5$ for silicate grains; see \citealt[and references therein]{2016ApJ...831..159H}). This equation shows the dependence on $T_{\rm d}$. The magnetic susceptibility reduces for higher $T_{\rm d}$ because of the halt in magnetization due to the thermal fluctuation of free-electron spins.

For a superparamagnetic grain, we get
\begin{equation}
    \chi_{\rm SPM}(0) = 0.026N_{\rm cl}\phi_{\rm sp}\hat{p}^{2}\left(\frac{20\,\rm K}{T_{\rm d}}\right)~{.}
\end{equation}
where $N_{\rm cl}$ is the number of Fe atoms per cluster, $\phi_{sp}$ is the volume filling factor of iron cluster in grain. $N_{\rm cl}$ could vary from 20 to $10^{5}$ (\citealt{1967ApJ...147..943J}). One can see that iron inclusion increases significantly the magnetic property of grains. 

For a grain rotating at a frequency $\omega$, the magnetic susceptibility is a complex number with the imaginary part that describes the absorption properties of magnetic energy given by
\begin{equation}
    \chi_{2}(\omega) = \frac{\chi(0)\tau_{\rm el}\omega}{[1+(\omega\tau_{\rm el}/2)^{2}]^{2}}~{,}
\end{equation}
where $\tau_{\rm el}=2.9\times 10^{-12}/f_{p}n_{23}$ is the spin-spin relaxation time with $n_{23}=n/10^{23}\,\rm cm^{-3}$ the atomic density of material, $f_{p}$ the fraction of Fe atoms in the grain (e.g., $f_{p}=1/7$ for MgFeSiO$_{4}$ grains, see \citealt{2014ApJ...790....6H}). 

Let us define a quantity $K(\omega)=\chi_{2}(\omega)/\omega$, which yields
\begin{equation}
    K_{\rm PM}(\omega) \simeq 8.7\times 10^{-14}\hat{p}\left(\frac{20\,\rm K}{T_{\rm d}}\right)\left(\frac{1}{[1+(\omega\tau_{\rm el}/2)^{2}]^{2}}\right)~{,}
\end{equation}
 
and
\begin{equation}
    K_{\rm SPM}(\omega) \simeq  2.6\times 10^{-11}N_{\rm cl}\phi_{\rm sp}\hat{p}^{2}\left(\frac{20\,\rm K}{T_{\rm d}}\right)k_{\rm SPM}(\omega)~{,}
\end{equation}
with
\begin{equation}
    k_{\rm SPM}(\omega)= \exp\left(\frac{{\rm 0.011\,K}\times N_{\rm cl}}{T_{\rm d}}\right)\left[1+\left(\frac{\omega\tau_{\rm sp}}{2}\right)^{2}\right]^{-2},\label{eq:gsp}
\end{equation}
where $\tau_{\rm sp}$ is the timescale of thermally activated remagnetization given by
\begin{equation}
    \tau_{\rm sp} \simeq 10^{-9}\exp\left(\frac{0.011\,{\rm K}\times N_{\rm cl}}{T_{\rm d}}\right)~{\rm s.}
\end{equation}

The supermagnetic grain has enormously increased magnetic susceptibility (\citealt{1967ApJ...147..943J}), and the coupling with the magnetic field is stronger than in an ordinary paramagnetic grain. Note that a superparamagnetic grain behaves like a paramagnetic grain in the absence of an external magnetic field. 

\subsection{Grain suprathermal rotation by RATs and rotational damping}\label{sec:St}
Let us consider a grain exposed to a radiation field of the energy density of $u_{\rm rad}\,(\rm erg\,cm^{-3})$, the mean wavelength of $\bar{\lambda}$ and an anisotropy degree of $\gamma$\footnote{$\gamma=0$ for a isotropic radiation field, and $\gamma=1$ for an unidirectional radiation field. For a typical diffuse interstellar radiation field: $\gamma\simeq 0.1$. For molecular cloud: $\gamma \simeq 0.7$ (\citealt{1996ApJ...470..551D})}. Its strength is defined by a dimensionless $U=u_{\rm rad}/u_{\rm ISRF}$, where $u_{\rm ISRF}=8.64\times 10^{-13}\,\rm erg\,cm^{-3}$ is the radiation energy density of the interstellar radiation field (ISRF) in the solar neighborhood (\citealt{1983A&A...128..212M}). 

The rotation of an irregular grain of size $a$ is described by the suprathermal number $St=\Omega/\Omega_{\rm T}$ with $\Omega_{\rm T}=(k_{\rm B}T_{\rm gas}/I_{\|})^{0.5}$ the thermal angular velocity. Grain rotation is called "suprathermal" for $St>1$ and "subthermal" for $St<1$. 

One of the important effects of RATs is to spin-up the dust grains of irregular shapes. On the another hand, gas collisions and IR re-emission spin-down the grain (see \citealt{2019ApJ...876...13H}). The timescale characterizing the grain rotational damping by gas collisions and IR emission is 
\begin{equation}
    \tau_{\rm damp} = \frac{\tau_{\rm gas}}{1+F_{\rm IR}}
\end{equation}
where $\tau_{\rm gas}$ and $F_{\rm IR}$ are the gas collision damping timescale and the ratio of the IR re-emission to collisional damping times, which are
\begin{equation}
    \tau_{\rm gas}=\frac{3}{4\sqrt{\pi}}\frac{I_{\|}}{1.2n_{\rm H}m_{\rm H}
v_{\rm T}a^{4}\Gamma_{\|}} \simeq 8.3\times 10^{3}\hat{\rho}\left(\frac{sa_{-5}}{n_{3}T_{1}^{1/2}\Gamma_{\|}}\right)~{\rm yr},\label{eq:tgas}
\end{equation}
\begin{equation}
    F_{\rm IR} \simeq 3.8\times 10^{-2}\left(\frac{U^{2/3}}{a_{-5}}\right)\left(\frac{10^{3}\, \rm{cm^{-3}}}{n_{\rm H}}\right)\left(\frac{10\,\rm K}{T_{\rm gas}}\right)^{1/2}.
\end{equation}
with $n_{3}=n_{\rm H}/(10^{3}\,\rm cm^{-3})$ the gas volume density, $T_{1}=T_{\rm gas}/10\,\rm K$ the gas temperature, $a_{-5}=a/(10^{-5}\,\rm cm)$ the grain size, $\hat{\rho}=\rho/(3\,\rm g\,cm^{-3})$ the mass density of the grain, and $\Gamma_{\|}$ the geometrical parameter ($\Gamma_{\|}=1$ for a spherical grain and $0.82$ for a grain of axial ratio $s=1/2$; see \citealt{Hoang.2022}).

The balance between the spin-up by RATs and spin-down by gas collisions and IR emission establishes the maximum rotation rate of grains due to RATs. 
For a simple case of a constant luminosity source\footnote{If the radiation source varies in luminosity, the angular velociy $\Omega_{\rm RAT}$ is obtained by solving the equation of motion; see Equation 12 in \cite{2019NatAs...3..766H}}, the maximum angular velocity that an irregular grain acquired by RATs is \begin{equation}
    \Omega_{\rm RAT} = \frac{\bar{\Gamma}_{\rm RAT}\tau_{\rm damp}}{I_{\parallel}}
\end{equation}
where $\bar{\Gamma}_{\rm RAT}$ is the average RAT acting on grain, which depends on the RAT efficiency ($Q_{\Gamma}$), grain size ($a_{\rm eff}$) and local radiation ($u_{\rm rad}$, $\bar{\lambda}$, $\gamma$) as
\begin{equation}
    \bar{\Gamma}_{\rm RAT} = \pi a_{\rm eff}^{2}\left(\frac{\bar{\lambda}}{2\pi}\right)\gamma u_{\rm rad}\bar{Q}_{\Gamma}
\end{equation}

The RAT efficiency strongly depends on the ratio of radiation wavelength and grain size as $\overline{Q}_{\Gamma} \propto (\lambda/a)^{-\eta}$\footnote{the power index $\eta$ describes the changes in RAT efficiency with respect to grain size; see Figure 1 in \citealt{2021ApJ...908..218H}} (see e.g., \citealt{2019NatAs...3..766H,2021ApJ...908..218H}), indicating that RATs could be effective at longer wavelengths for large grains (e.g., optical-IR photons) rather than only UV photons. Thus, the RAT effect is effective in more extended regions than other dust effects induced by UV photons, such as photo-electron emission. 

Because the RAT efficiency changes its slope at $a\simeq \bar{\lambda}/2.5$ (see e.g., \citealt{2021ApJ...908..218H}), the suprathermal number of grains spun-up by RATs is described separately as (\cite{Hoang.2022}) 
\begin{equation}
St_{\rm RAT} \simeq 180 \hat{\rho}^{1/2}s^{5/6}a_{-5}^{7/2}\left(\frac{\bar{\lambda}}{1.2\,\rm \mu m}\right)^{-2}
\left(\frac{\gamma U}{n_{3}T_{1}}\right)\left(\frac{1}{1+F_{\rm IR}}\right),\label{eq:S_RAT1}
\end{equation}
for $a\lesssim \bar{\lambda}/2.5$, and
\begin{equation}
St_{\rm RAT}\simeq 2.1\times 10^{4}\hat{\rho}^{1/2}s^{-1/6}a_{-5}^{1/2}\left(\frac{\bar{\lambda}}{1.2\,\rm \mu m}\right) \left(\frac{\gamma U}{n_{3}T_{1}}\right)\left(\frac{1}{1+F_{\rm IR}}\right).\label{eq:S_RAT2}
\end{equation}
for $a>\bar{\lambda}/2.5$.

For regions where the gas density is very high and the radiation strength is rather small (i.e., $F_{\rm IR}\ll 1$), such as in pretostellar cores and the mid-plane of protoplanetary disks, one can see that $St_{\rm RAT}\propto U/n_{3}T_{1}$. For regions with very strong radiation fields but a low gas density (i.e., $F_{\rm IR}\gg 1$), such as in the vicinity of a radiation source, one has $St_{\rm RAT}\propto U^{1/3}$. Note that $St_{\rm RAT}$ also strongly depends on the size of the grains, the bigger grain size, the faster rotation induced by RATs. Therefore, large grains are easily to be spun-up suprathermally in strong radiation fields.

\subsection{Internal alignment}\label{sec:internal}
The internal alignment describes the alignment of the angular momentum $\boldsymbol{J}$ with the short axis $\boldsymbol{a_{1}}$ of the grain as illustrated in Figure \ref{fig:grain_define}(a). Here we discuss three relaxation processes that induce internal alignment, including Barnett relaxation, nuclear relaxation, and inelastic relaxation, as well as show the critical grain sizes for each internal alignment.

\subsubsection{Barnett relaxation}
The most important process relies on the Barnett relaxation (\citealt{1979ApJ...231..404P}). A rotating (super-)paramagnetic grain will be magnetized due to the Barnett effect, which results in the dissipation of rotational energy due to the rotating component perpendicular to $\boldsymbol{a_{1}}$. The dissipation will eventually bring $\boldsymbol{\Omega}$ and $\boldsymbol{J}$ aligned with $\boldsymbol{a_{1}}$, which is the minimum energy state. The stage of $\boldsymbol{J}\perp \boldsymbol{a_{1}}$ is called "right internal alignment".

The relaxation times by the Barnett effect for a paramagnetic (so-called Barnett relaxation) and a supermagnetic grain (so-called super-Barnett relaxation) are given by
\begin{equation}
    \tau_{\rm BR,PM}=\frac{\gamma_{e}^{2}I_{\|}^{3}}{VK_{\rm PM}(\omega)h^{2}(h-1)J^{2}} \simeq 0.5\hat{\rho}^{2}a^{7}_{-5}f(\hat{s})\left(\frac{J_{\rm d}}{J}\right)^{2}\left[1+\left(\frac{\omega \tau_{\rm el}}{2}\right)^{2}\right]^{2}~{\rm yr,}
\end{equation}
\begin{equation}
    \tau_{\rm BR,SPM}=\frac{\gamma_{e}^{2}I_{\|}^{3}}{VK_{\rm SPM}(\omega)h^{2}(h-1)J^{2}}
\approx 0.16 \hat{\rho}^{2}f(\hat{s})a_{-5}^{7}\left(\frac{1}{N_{\rm cl}\phi_{\rm sp,-2}\hat{p}^{2}}\right)\left(\frac{J_{\rm d}}{J}\right)^{2}\left(\frac{T_{d,1}}{k_{\rm SPM}(\omega)}\right)~ \rm yr,
\label{eq:tauBar_sup}\\
\end{equation}
where $V=4\pi/3 a^{3}$ is the grain volume, $\gamma_{e}=-g_{e}\mu_{B}/\hbar$ is gyromagnetic ratio of an electron ($g_{e}\simeq 2$ and $\mu_{B}\simeq 9.26\times 10^{-21}\,\rm erg\, G^{-1}$), $\hat{s}=s/0.5, f(\hat{s})=\hat{s}[(1+\hat{s}^{2})/2]^{2}$, and $J_{\rm d}=\sqrt{I_{\|}k_{\rm B}T_{\rm d}/(h-1)}$ is the dust thermal angular momentum (see also \citealt{2014MNRAS.438..680H}) with $h=I_{\|}/I_{\perp}=2/(1+s^{2})$ and $\phi_{\rm sp,-2}=\phi_{\rm sp}/0.01$. 

However, gas collisions can strongly randomize the internal alignment if the Barnett relaxation is slower than the grain randomization by gas collisions. Therefore, grains have efficient internal alignment when the timescale for the Barnett relaxation is shorter than that for gas rotational damping. Thus, the maximum size that a paramagnetic and superparamagnetic grain has the internal alignment with $\boldsymbol{J}$ along the short axis $\boldsymbol{a_{1}}$ can be determined by the balance between these two timescales, which yields (\cite{Hoang.2022})
\begin{equation}
a^{\rm PM}_{\rm max,aJ} \simeq 0.39\times h^{1/3}St^{1/3} \left(\frac{\hat{p}^{2}}{\hat{\rho}n_{3}T^{1/2}_{1}}\right)^{1/6}
\times \left(\frac{1}{1+(\omega \tau_{\rm tel}/2)^{2}}\right)^{1/3}\left(\frac{(h-1)T_{\rm gas}}{T_{\rm d}}\right)^{1/6}~{\rm \mu m},
\end{equation}
and
\begin{equation}
a^{\rm SPM}_{\rm max,aJ}\simeq 2.18h^{1/3}St^{1/3}\left(\frac{N_{\rm cl,4}\phi_{\rm sp,-2}\hat{p}^{2}}{\hat{\rho}n_{3}T_{1}^{1/2}\Gamma_{\|}}\right)^{1/6}
\times\left(\frac{1}{k_{\rm SPM}(\omega)}\right)^{1/6}\nonumber\\
\times\left(\frac{(h-1)T_{\rm gas}}{T_{\rm d}}\right)^{1/6}~{\rm \mu m},\label{eq:amax_aJ_BR}
\end{equation}
with $N_{\rm cl,4}=N_{\rm cl}/10^{4}$ \citep{Hoang.2022}. One can see that $a_{\rm max,aJ}$ increases with grain rotation as $St^{1/3}$ and decreases with gas density as $\sim n^{-1/6}_{\rm H}$. With iron inclusion, superparamagnetic grains could achieve efficient internal alignment for larger grain sizes.

\subsubsection{Nuclear relaxation}
Atomic nuclei within the grain can also have {\it nuclear} paramagnetism due to unpaired protons and nucleons. The attachment of H atoms to the grain or water ice mantle induce nuclear magnetism due to proton spin (\citealt{1979ApJ...231..404P}). Nuclear paramagnetism can also induce internal relaxation as Barnett effect for electron spins (\citealt{1999ApJ...520L..67L}). The nuclear relaxation (NR) time is given by
\begin{eqnarray}
    \tau_{\rm NR}&=&\frac{\gamma_{n}^{2}I_{\|}^{3}}{VK_{n}(\omega)h^{2}(h-1)J^{2}},\nonumber\\
&\simeq&125\hat{\rho}^{2}a_{-5}^{7}f(\hat{s})\left(\frac{n_e}{n_n}\right) \left(\frac{J_{d}}{J}\right)^{2}
\left(\frac{g_n}{3.1}\right)^{2}\left(\frac{2.79\mu_N}{\mu_n}\right)^{2}\times \left[1+\left(\frac{\omega\tau_{n}}{2}\right)^{2}\right]^{2}{\rm s},
\label{eq:tau_nucl}
\end{eqnarray}
where $\gamma_{n}$ is nuclear gyromagnetic ratio, $n_{e}$ is the number density of unpaired electrons in the dust, $n_{n}$ is the number density of nuclei that has the magnetic moment $\mu_{n}$, and $\mu_{N}=e\hbar/2m_{p}c=5.05 \times10^{-24}$ erg G$^{-1}$ with $m_{p}$ the proton mass is the nuclear magneton, $\tau_{n}\sim 10^{-5}(3.1/g_{n})^{2}(10^{22}{\rm cm}^{-3}/n_{e})$s is the nuclear spin-spin relaxation time, and $K_{n}=\chi_{n}/\omega$ with $\chi_{n}$ the nuclear magnetic susceptibility (see \citealt{2022ApJ...928..102H}). 

The maximum size for internal alignment induced by nuclear relaxation is given by \cite{2022ApJ...928..102H},
\begin{eqnarray}
    a_{\rm max,aJ}({\rm NR}) < 3.03 h^{1/3}St^{1/3}\left(\frac{(n_{e}/n_{n})(g_{n}/3.1)^{2}}{n_{3}T_{1}^{1/2}\Gamma_{\|}}\right)^{1/6}\left(\frac{1}{1+(\omega\tau_{\rm n}/2)^{2}}\right)^{1/3} \times\left(\frac{(h-1)T_{\rm gas}}{T_{\rm d}}\right)^{1/6}~{\rm \mu m.}\label{eq:a_nucl}
\end{eqnarray}
which increases with the suprathermal rotation as $St^{1/3}$ for $\omega<2/\tau_{n}$, but decreases as $St^{-1/3}$ for $\omega>2/\tau_{n}$ due to suppression of nuclear magnetism at fast rotation. For the typical density of $n=10^{3}\,\rm cm^{-3}$, large grains of size upto $a\sim 3\,\mu$m can have efficient internal alignment by nuclear relaxation.

\subsubsection{Inelastic relaxation}\label{sec:inelastic}
Atoms and molecules in a rotating grain experience a centrifugal force that stress them out from the grain central of mass. The chemical attractive forces, on the other hand, tend to tight them together. These effects cause the grain deformation. If grains are made up inelastic materials, the deformation results in the dissipation of rotational energy, which eventually brings the angular momentum $\boldsymbol{J}$ to align with the grain's shortest axis $\boldsymbol{a_{1}}$ (see \citealt{1979ApJ...231..404P,1999MNRAS.303..673L}). This internal alignment mechanism is named as inelastic relaxation.

The characteristic time of inelastic relaxation for an oblate spheroidal grain can be estimated as (see \citealt{Hoang.2022})
\begin{equation}
\tau_{\rm iER}= \frac{\mu Q}{\rho a^{2}\Omega_{0}^{3}}g(s)=\frac{\rho^{1/2}a^{11/2}\mu Q }{(kT_{\rm gas})^{3/2}St^{3}}g'(s)=0.034\hat{\rho}^{1/2}a_{-5}^{11/2}\frac{\mu_{8}Q_{3}}{T_{1}^{3/2}}\frac{g'(s)}{St^{3}}~{\rm yr},\label{eq:ti_LE}
\end{equation}
where $Q$ is the quality factor of grain material (Q=100 for a silicate rocks, \citealt{Efroimsky:2000p5384}), $\mu$ is the modulus of rigidity ($\mu\sim 10^{7}\,\rm erg\,cm^{-3}$ for cometary, \citealt{Knapmeyer.2018}). $\Omega_{0}=St \Omega_{T}$, $\mu_{8}=\mu/(10^{8}\,\rm erg\,cm^{--3})$, $Q_{3}=Q/10^{3}$, and $g'=2.2s^{3/2}g(s)$ with $g(s)$ the geometrical factor as
\begin{equation}
g(s)=\frac{ 2^{3/2}7}{8}\frac{(1+s^{2})^{4}}{s^{4}+1/(1+\sigma)},\label{eq:gs}
\end{equation}
which corresponds to $g(s)=7.0$ and $4.6$ for $s=1/2$ and $1/3$, respectively.

Following \cite{Hoang.2022}, the maximum grain size for efficient inelastic relaxation is defined by $\tau_{\rm iER}=\tau_{\rm gas}$ and given by
\begin{equation}
    a_{\rm max, aJ}({\rm iER})\simeq 1.55\left(\frac{\mu_{8}Q_{3}}{\sqrt{\hat{\rho}}}\right)^{-2/9}\left(\frac{T_{1}}{n_{3}}\right)^{2/9}
\times\left(\frac{s}{g'(s)\Gamma_{\|}}\right)^{2/9}St^{2/3}~{\rm \mu m,}\label{eq:amax_aJ_iER}
\end{equation}
which implies that large grains of $a\sim 1.55\,\rm \mu m$ can be efficiently aligned via inelastic relaxation even at thermal rotation of $St=1$, and large grains can be aligned for $St>1$.

\begin{figure}
    \centering
    \includegraphics[width=0.95\textwidth]{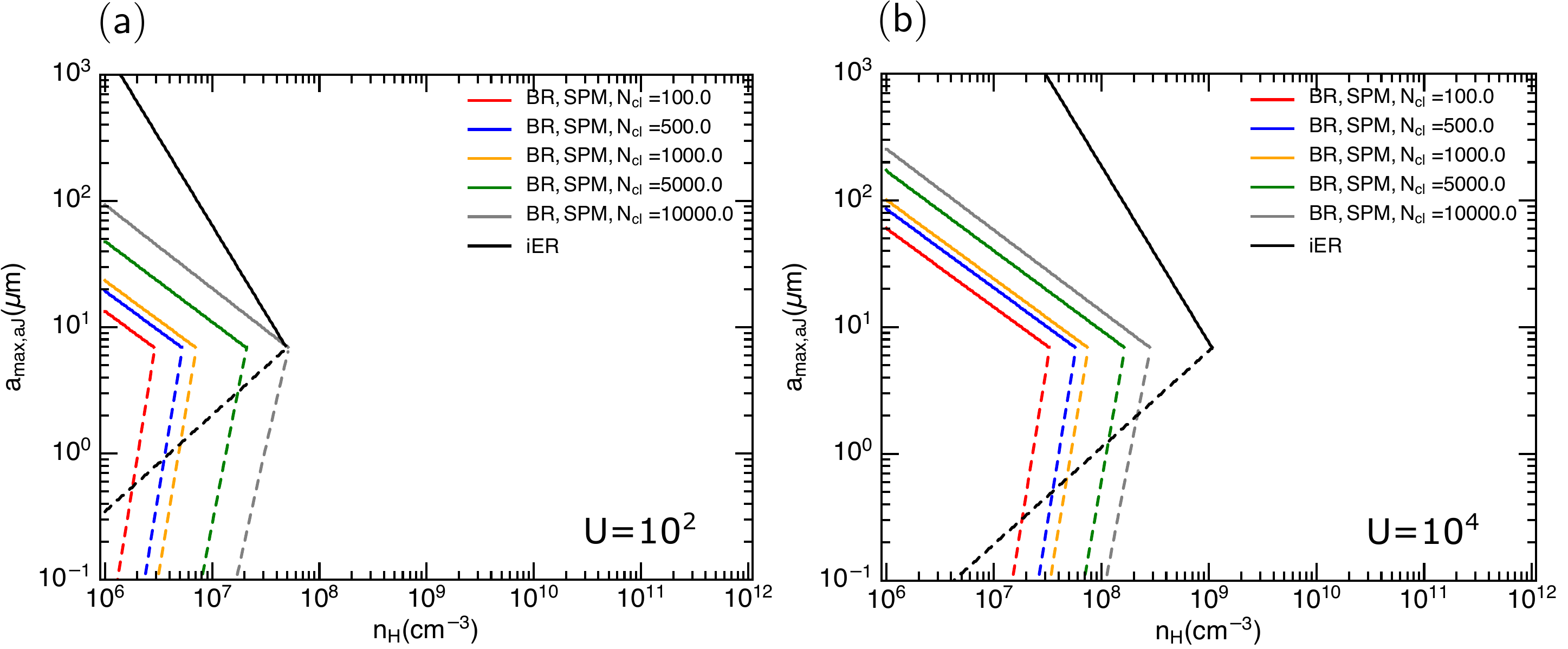}
    \caption{Comparison of minimum (dashed line) and maximum (solid line) size of grains having internal alignment by Barnett relaxation (BR) for superparamagnetic grain (SPM, color lines) and inelastic mechanism (iER, black lines) with respect to a large range of gas density (Figure adopted from \citealt{Hoang.2022}). The grains are spun-up by RATs with $U=10^{2}$ (panel (a)) and $U=10^{4}$ (panel (b)).}
    \label{fig:IE}
\end{figure}
Figure \ref{fig:IE} shows the comparison of grains having internal alignment by Barnett relaxation for superparamagnetic grains (color lines) and inelastic relaxation (black lines) as functions of the gas density. Grains are assumed to be spun-up by RATs. First, the Barnett relaxation is not effective in very dense regions (e.g., $n_{\rm H}>10^{9}\,\rm cm^{-3}$) because of the gas rotational damping and lower $St_{\rm RAT}$. For a given gas density, one can see that Barnett relaxation is more efficient, but the inelastic relaxation is able to align larger grains (panel (a)). The reason is that the maximum grain size for inelastic relaxation scales as $\sim St^{2/3}$, while it scales as $\sim St^{1/3}$ for Barnett relaxation. For stronger radiation fields, the inelastic relaxation is efficient in denser gas (panel (b)), which is due to the fact that inelastic timescale reduces as $\sim St^{-1/3}$. 

For a typical density of the ISM ($n_{\rm H}\simeq 10^{3}\,\rm cm^{-3}$), one can see that the Barnett relaxation for PM/SPM grains are much more effective than the inelastic relaxation. However, unlike the PM/SPM, diamagnetic grains (e.g., carbonaceous dust) does not experience the Barnett relaxation effect, but they could still have efficient internal alignment for size $a<a_{\rm max,aJ}(\rm iER)$ due to inelastic relaxation, because this relaxation is independent from the grain magnetic properties.

\subsection{External alignment}
The external alignment describes the alignment of the grain angular momentum $\boldsymbol{J}$ with a preferred direction in space, which can be the ambient magnetic field or the radiation direction (see Figure \ref{fig:grain_define}(b)).

\subsubsection{Alignment with magnetic fields (magnetic alignment): Larmor precession and magnetic relaxation}

{\bf Larmor precession}: The interaction between the grain magnetic moment due to the Barnett effect with the magnetic field induces the Larmor precession of $\boldsymbol{J}$ around $\boldsymbol{B}$, resulting the first coupling of the grain with the magnetic field (see e.g., \citealt{2022ApJ...928..102H}).

The rate of the Larmor precession for (super)paramagnetic grains are respectively given by
\begin{equation}
\tau_{B,\rm PM}=\frac{2\pi}{|d\phi/dt|}=\frac{2\pi I_{\|}\Omega}{|\mu_{\rm Bar}|B}=\frac{2\pi |\gamma_{e}|I_{\|}}{\chi_{\rm PM}(0)VB}\simeq 3.5\times 10^{-4}\frac{\hat{\rho}T_{d,1}a^{2}_{-5}}{n_{23}f_{p}\hat{p}^{2}}\left(\frac{B}{20\,\rm \mu G}\right)^{-1}~{\rm yr,}  
\end{equation}
and
\begin{equation}
\tau_{B,\rm SPM}=\frac{2\pi |\gamma_{e}|I_{\|}}{\chi_{\rm SPM}(0)VB}\simeq
4.05\times 10^{-2}\frac{\hat{\rho} T_{d,1}a_{-5}^{2}}{N_{\rm cl}\phi_{sp,-2}\hat{p}^{2}}\left(\frac{B}{20\,\rm \mu G}\right)^{-1} \rm yr.~~~~
\label{eq:tauB}
\end{equation}

When the Larmor precession is faster than the gas randomization, the magnetic field becomes an axis of grain alignment, which is known as {\it magnetic alignment}. The maximum size for the fast Larmor precession around the B-field is defined by $\tau_{B}=\tau_{\rm gas}$, which is
\begin{equation}
    a^{\rm Lar,PM}_{\rm max,JB} \simeq 1.2\times10^{6} \hat{s}n^{-1}_{3}T^{-1/2}_{1}T^{-1}_{d,1}\left(\frac{B}{20\,\rm \mu G}\right)\left(\frac{n_{23}f_{p}\hat{p}^{2}}{\Gamma_{\|}}\right)~{\rm \mu m,}
\end{equation}
and
\begin{equation}
    a^{\rm Lar, SPM}_{\rm max, JB} \simeq 1.02\times 10^{9} \hat{s}n^{-1}_{3}T_{1}^{-1/2}T^{-1}_{d,1} \left(\frac{B}{20\,\rm \mu G}\right)\left(\frac{N_{\rm cl,4}\phi_{\rm sp,-2}\hat{p}^{2}}{\Gamma_{\|}}\right)~{\rm \mu m.}~~~\label{eq:amax_JB}
\end{equation}
for PM and SPM grains, respectively.

{\bf Para- and superpara-magnetic relaxation}: The angle between $\boldsymbol{J}$ and the magnetic field can be reduced by paramagnetic relaxation, resulting in the magnetic alignment of $\boldsymbol{J}$ with $\boldsymbol{B}$. This mechanism is known as the DG relaxation (\citealt{1951ApJ...114..206D}). 
As numerically shown in \cite{2016ApJ...831..159H,2016ApJ...821...91H}, magnetic relaxation alone is not enough to produce efficient grain alignment when grains have thermal rotation (i.e., $St\leq 1$) due to strong internal thermal fluctuations and gas randomization. Yet, the joint action of suprathermal rotation by RATs and magnetic relaxation however can make grains to achieve high degree of alignment.

The characteristic times of the DG relaxation for the PM and SPM grains are given by (see e.g., \citealt{2016ApJ...831..159H}) 
\begin{equation}
\tau_{\rm mag,PM} = \frac{I_{\|}}{VK_{\rm PM}(\Omega)B^{2}}=\frac{2\rho a^{2}}{5K_{\rm PM}(\Omega)B^{2}}
\simeq 8.6\times 10^{4}\frac{\hat{\rho}a^{2}_{-5}T_{d,1}}{\hat{p}}\left(\frac{B}{20\,\rm \mu G}\right)^{-2}\left(1+(\omega \tau_{\rm el}/2)^{2}\right)^{2} ~{\rm yr,}\label{eq:tau_DG}~~~
\end{equation}
and
\begin{equation}
\tau_{\rm mag,SPM} = \frac{2\rho a^{2}}{5K_{\rm SPM}(\Omega)B^{2}}
\simeq 
3.75\times 10^{2}\frac{\hat{\rho}a_{-5}^{2}}{N_{\rm cl}\phi_{\rm sp}\hat{p}^{2}}\frac{T_{d,1}}{k_{\rm SPM}(\Omega)}\left(\frac{B}{20\,\rm \mu G}\right)^{-2} ~{\rm yr.}\label{eq:tau_DG}~~~
\end{equation}

The respective maximum of size that grain alignment can still be affected by the DG magnetic relaxation is defined by $\tau_{\rm mag}=\tau_{\rm gas}$ and is equal to \citep{Hoang.2022}
\begin{equation} \label{eq:a_DG_pm}
    a^{DG,\rm PM}_{\rm max,JB} \simeq 0.0097\times n^{-1}_{3} T^{-1/2}_{1}T^{-1}_{d,1}
    \left(\frac{B}{20\,\rm \mu G}\right)^{2}\left(1+(\omega \tau_{el}/2)^{2}\right)^{-2}~{\rm \mu m,}
\end{equation}
and
\begin{equation}
    a_{\rm max,JB}^{DG,\rm SPM}\approx 224\times n^{-1}_{3} T^{-1/2}_{1}T^{-1}_{d,1} \left(\frac{B}{20\,\rm \mu G}\right)^{2}N_{\rm cl,4}\phi_{\rm sp,-2}\hat{p}^{2}k_{\rm SPM}(\omega)~{\rm \mu m.}\label{eq:amax_JB_DG}
\end{equation}
One can see that the gas density and B-field strength determine which grains are aligned with B-field as $a^{DG}_{\rm max, JB}\sim n^{-1}_{\rm H} B^{2}$. The DG efficiency drops dramatically with increasing the gas density. Paramagnetic relaxation is negligible for large grains, but superparamagnetic relaxation can be efficient even for large grains up to $\sim 20\,\rm \mu m$ for MCs with $n_{\rm H}\sim 10^{3}\,\rm cm^{-3}$.

\subsubsection{The RAT Alignment Paradigm}\label{sec:highJ_attractor}
Here we describe the basic elements of the RAT alignment paradigm, including radiative precession, grain alignemnt at low-$J$ and high-$J$ attractors, and the minimum size for RAT alignment.

{\bf Radiative precession}: In the presence of an anisotropic radiation field, irregular grains experience radiative precession due to RATs, with the grain angular momentum precessing around the radiation direction ($\boldsymbol{k}$). For a grain with angular momentum $J$, the radiative precession time is given by (\citealt{2007MNRAS.378..910L}; \citealt{2014MNRAS.438..680H}),
\begin{equation}
\tau_{k}=\frac{2\pi}{|d\phi/dt|}\approx\frac{2\pi J}{\gamma u_{\rm rad}\lambda a_{\rm eff}^{2}Q_{e3}}
\simeq 56.8
\hat{\rho}^{1/2}T_{1}^{1/2}\hat{s}^{-1/6}a_{-5}^{1/2}St\left(\frac{1.2\,\rm \mu m}{\gamma\bar{\lambda}\hat{Q}_{e3}U}\right)
\rm yr,~~~\label{eq:tauk}
\end{equation} 
where $\hat{Q}_{e3}=Q_{e3}/10^{-2}$ with $Q_{e3}$ the third component of RATs that induces the grain precession around $\boldsymbol{k}$ and the normalization is done using the typical value of $Q_{e3}$ (see \citealt{2007MNRAS.378..910L}).  

In the absence of magnetic field, radiative precession can be much faster than gas damping time for a thermal rotation of $St\sim 1$ (Eq. \ref{eq:tgas}). In this case, the anisotropic radiation direction is the axis of grain alignment, which is known as k-RAT alignment. 

{\bf Low$-J$ and high$-J$ Attractor Alignment by RATs}: In addition to radiative precession, RATs can align grains at high$-J$ attractors and low$-J$ attractors. In general, a fraction of grains ($f_{\rm high-J}$) can be aligned at a high$-J$ attractor due to the aligning and spin-up effects of RATs (\citealt{2007MNRAS.378..910L,2008MNRAS.388..117H}). Grains aligned at the high$-J$ attractor have the suprathermal rotation with $St_{\rm high-J}>1$. On the other hand, due to the aligning and spin-down effects of RATs, the $1-f_{\rm high-J}$ fraction of grains, are aligned at the low$-J$ attractor with $St_{\rm low-J}\sim 1$ \citep{2014MNRAS.438..680H}. The RAT alignment process can occur on a timescale smaller than the gas damping time, which is called {\it fast} alignment \citep{2007MNRAS.378..910L,2021ApJ...908...12L} with a strong radiation field. Grains can be stably aligned at low$-J$ attractors when the grain randomization by gas collisions is not considered. Otherwise, the gas randomization could randomize grains at low$-J$ attractors and eventually transport them to the high$-J$ attractor after a timescale greater than the gas damping time, which is usually called {\it slow} alignment \citep{2008MNRAS.388..117H,2021ApJ...908...12L}. The {\it fast} RAT alignment occurs in the local environment of intense transient radiation sources such as supernova \citep{2020ApJ...888...93G}, kilonova, and gamma-ray bursts \citep{2020ApJ...895...16H}, whereas the {\it slow} RAT alignment occurs in most astrophysical environments, from the ISM, MCs, to SFRs. A detailed discussion of fast and slow alignment by RATs is discussed in detail in \cite{2021ApJ...908...12L}.

The exact value of $f_{\rm high-J}$ depends on the grain properties (shape and size) and magnetic susceptibility.  For grains with ordinary paramagnetic material (e.g., silicate), \cite{2021ApJ...913...63H} found that $f_{\rm high-J}$ can be about $10-70\%$ based on calculations of RATs for an ensemble of Gaussian random shapes.
The presence of iron inclusions embedded in the grains increases grain magnetic suceptibility and superparamagnetic relaxation, which can produce in the universal high$-J$ attractors(i.e., $f_{\rm high-J}\sim 100\%$) \citep{2016ApJ...831..159H,2021ApJ...908...12L}.

{\bf Minimum size of RAT alignment}: Due to gas collisions, grains can be efficiently aligned only when they rotate suprathermally at a high$-J$ attractor. \cite{2008MNRAS.388..117H} and \cite{2014MNRAS.438..680H} demonstrated that grain alignment is stablized under the gas randomization when $\Omega_{\rm RAT}\equiv 3\times \Omega_{\rm T}$ (or $St_{\rm RAT}=3$). This criteria yields the minimum size for grain alignment by RATs as
\begin{equation} \label{eq:a_align}
    a_{\rm align} \simeq 0.055\hat{\rho}^{-1/7} \left(\frac{n_{3} T_{1}}{\gamma_{-1} U}\right)^{2/7}
    \times \left(\frac{\bar{\lambda}}{1.2\,\rm \mu m}\right)^{4/7} \left(\frac{1}{1+F_{\rm IR}}\right)^{-2/7} ~\rm{\mu m},
\end{equation}
where $\gamma_{-1}=\gamma/10^{-1}$. One can see that for a typical MC with $n_{\rm H}=10^{3}\,\rm cm^{-3}$, grains larger than $a_{\rm align}=0.055\,\mu$m can be aligned by RATs. However, the radiation strength is not an observable quantity, thus we could use the dust temperature $T_{\rm d}$ as a proxy for this strength. If we adopt $U\simeq (T_{\rm d}/16.4\,\rm K)^{6}(a/10^{-5}\,\rm cm)^{6/15}$ (\citealt{2011piim.book.....D}), $a_{\rm align} \sim n^{2/7}_{\rm H}T^{-12/7}_{\rm dust}$ that illustrates that higher $T_{\rm d}$ is able to bring smaller grains to align. Vice-versa, only larger grains are able to align in higher $n_{\rm H}$ (denser gas). Note that the power-index in Equation \ref{eq:a_align} slightly differs from \cite{2021ApJ...908..218H} and \cite{2021ApJ...923..130T}. The reason is due to the slightly difference in the scaling of the average RAT efficiency, but this difference results in a negligible discrepancy of $a_{\rm align}$ (see Figure 1 in \citealt{2021ApJ...908..218H}).

{\bf k-RAT vs. B-RAT Alignment}: Using Equation (\ref{eq:tauk}) one calculates the radiative precession (k-RAT) timescale for grains aligned at low$-J$ attractors with $St=1$ as 
\begin{eqnarray}
    \tau_{k}^{\rm low-J}=56.4\hat{\rho}^{1/2}T_{1}^{1/2}\hat{s}^{-1/6}a_{-5}^{1/2}\left(\frac{1.2\mu m}{\gamma\bar{\lambda}\hat{Q}_{e3}U}\right)
{\rm yr},~~~~~\label{eq:tauk_lowJ}
\end{eqnarray}
and for grains aligned at high$-J$ attractors with $St=St_{\rm RAT}$ from Equation (\ref{eq:S_RAT2}), one has
\begin{eqnarray}
\tau_{k}^{\rm high-J}&=&1.17\times 10^{6}\hat{s}^{-1/3}\left(\frac{\hat{\rho}a_{-5}}{\hat{Q}_{e3}}\right)\left(\frac{1}{n_{3}T_{1}^{1/2}}\right)\left(\frac{1}{1+F_{\rm IR}}\right){\rm yr}.~~~~~\label{eq:tauk_highJ}
\end{eqnarray}

In the presence of a magnetic field, the grain experiences simultaneous Larmor precession and radiative precession. If the Larmor precession is faster than the radiative precession, i.e., $\tau_{B}<\tau_{k}$, the axis of grain alignment by RATs changes from with $\boldsymbol{J}$ along $\boldsymbol{k}$ to the magnetic field, $\boldsymbol{B}$, which is known as B-RAT. The minimum grain size that grain alignment still occurs via k-RAT, which is also the maximum size grains can be aligned via B-RAT, is defined by $\tau_{k}=\tau_{B}$ and is derived in \cite{Hoang.2022}. For grains aligned at a low$-J$ attractor, one has
\begin{equation}
    a^{\rm low-J,PM}_{\rm min,Jk}\equiv a_{\rm max,JB}^{
\rm low-J,PM} \simeq 2.96\hat{\rho}^{-1/2}\hat{s}^{-1/9} \left(\frac{\bar{\lambda}}{1.2\,\rm \mu m}\right)^{-2/3} 
    \left(\frac{n_{23}f_{\rm p}\hat{p}^{2}T^{1/2}_{1}}{T_{d,1}\gamma U_{3} \hat{Q}_{e3}}\right)^{2/3}\left(\frac{B}{20\,\rm \mu G}\right)^{2/3} ~{\rm \mu m,}
\end{equation}
and
\begin{eqnarray}
    a^{\rm low-J,SPM}_{\rm min,Jk}\equiv a_{\rm max,JB}^{\rm low-J,SPM} &=&57.8 \hat{\rho}^{-1/2}\hat{s}^{-1/9} \left(\frac{\bar{\lambda}}{1.2\,\rm \mu m}\right)^{-2/3}
    \left(\frac{\hat{p}^{2}T^{1/2}_{1}}{T_{d,1}\gamma_{\rm rad} U_{3}\hat{Q}_{e3}}\right)^{2/3} \\\nonumber
    &\times&\left(\frac{B}{20\,\rm \mu G}\right)^{2/3}
    \left(N_{\rm cl,4}\phi_{sp,-2}\right)^{2/3}~{\rm \mu m.}~~~~~\label{eq:amin_Jk_lowJ}
\end{eqnarray}
for PM and SPM grains, respectively. The above equations imply that the minimum size for k-RAT when grains aligned at low$-J$ attractors is smaller for higher radiation strength as $U^{-2/3}$, but larger for stronger magnetic field as $B^{2/3}$. It is harder to bring SPM grains to align with the radiation direction because of their larger magnetic susceptibility and then faster Larmor precession.

Similarly, for grains aligned at a high$-J$ attractor \citep{Hoang.2022}
\begin{equation}
    a^{\rm high-J,PM}_{\rm min,Jk} \equiv a_{\rm max,JB}^{\rm high-J,PM} \simeq 3.3\times 10^{8} \hat{s}^{-1/3}\left(\frac{n_{23}f_{\rm p}\hat{p}^{2}}{n_{3}T_{d,1}T^{1/2}_{1}\hat{Q}_{e3}}\right)
    \left(\frac{B}{20\,\rm \mu G}\right)\left(\frac{1}{1+F_{\rm IR}}\right) ~{\rm \mu m,}
\end{equation}
and
\begin{equation}
    a^{\rm high-J,SPM}_{\rm min,Jk}\equiv a_{\rm max,JB}^{\rm high-J,SPM} \simeq 2.8\times 10^{10}\hat{s}^{-1/3}\left(\frac{\hat{p}^{2}}{n_{3}T_{d,1}T_{1}^{1/2}\hat{Q}_{e3}}\right) 
    \left(\frac{B}{20\,\rm \mu G}\right)
    \left(\frac{N_{\rm cl,4}\phi_{sp,-2}}{1+F_{\rm IR}}\right)~{\rm \mu m.}~~~\label{eq:amin_Jk_highJ}
\end{equation}
One realizes that the minimum size for k-RAT alignment is very large for typical molecular clouds, especially when grains are aligned at high$-J$ attractors. Therefore, in these conditions, grains are aligned with $\boldsymbol{J}$ along the magnetic field via B-RAT. Yet, for very strong radiation fields of strength of $U\gg 1$, small grains at low$-J$ attractors can have k-RAT alignment.

\subsection{Radiative Torque Disruption (RAT-D)}
\subsubsection{The RAT-D mechanism}
A dust grain rotating with an angular velocity $\Omega$ develops a tensile stress $S=\rho \Omega^{2}a^{2}/4$ on grain material. This stress points outward from the grain center of mass and tends to tear the grain apart. When the tensile stress exceeds the maximum tensile strength of the grain material, $S_{\rm max}$, the grain is instantaneously disrupted in fragments. The maximum angular velocity beyond which the grain is disrupted is defined by $S=S_{\rm max}$ and reads
\begin{equation} \label{eq:omega_crit}
    \Omega_{\rm disr} = \frac{2}{a}\left(\frac{S_{\rm max}}{\rho}\right)^{1/2} \simeq \frac{3.65\times 10^{8}}{a_{-5}}S^{1/2}_{\rm max,7}\hat{\rho}^{-1/2} {~\rm rad/s},
\end{equation}
where $S_{\rm max,7}=S_{\rm max}/(10^{7} \rm erg\,cm^{-3})$. The exact value of $S_{\rm max}$ depends on the grain internal structure (see \citealt[and references therein]{2020Galax...8...52H}). For example, composite grains have $S_{\rm max} \sim 10^{7}\,\rm erg\,cm^{-3}$, whereas compact grains have a higher value.

\begin{figure}
    \centering
    \includegraphics[width=0.7\textwidth]{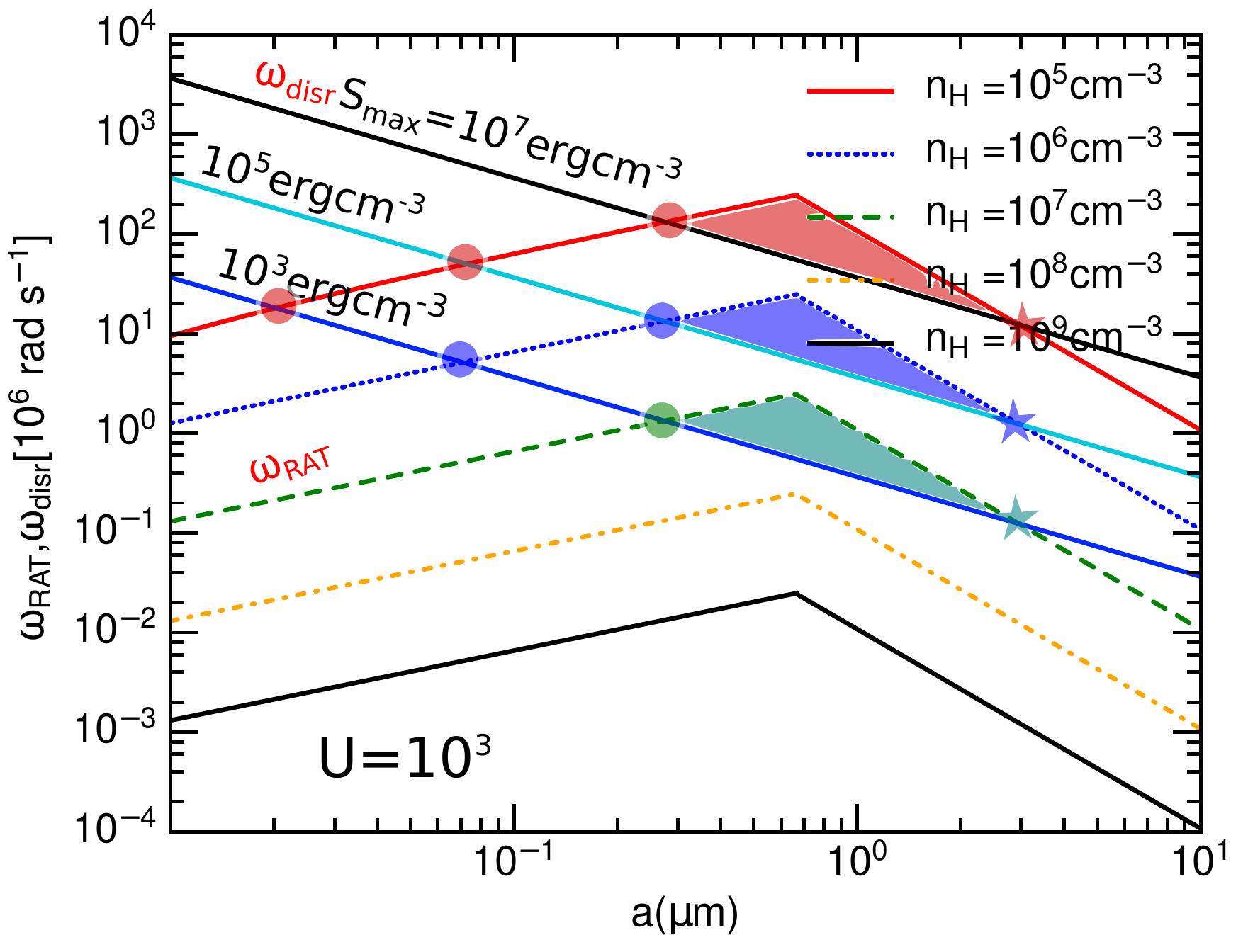}
    \caption{Illustration of the rotation rate by RATs and disruption limit for different tensile strength. The lower and upper intersections define the lower and upper disruption sizes. Figure adopted from \cite{2020Galax...8...52H}.}
    \label{fig:omega_rat}
\end{figure}

Subject to a strong radiation field, dust grains will be spontaneously fragmented into many smaller species when the grain angular velocity spun-up by RATs, $\Omega_{\rm RAT}$, exceeds the critical value $\Omega_{\rm disr}$. This is called the radiative torque disruption (RAT-D) mechanism, which was first introduced in \cite{2019NatAs...3..766H}. Due to the dependence of $\Omega_{\rm RAT}$ on the grain size and local conditions, there is a range of grain sizes that can be disrupted, as illustrated by the shaded areas in Figure \ref{fig:omega_rat}. The minimum size of grains that can be disrupted by RAT-D is
\begin{eqnarray}
a_{\rm disr}&=&\left(\frac{0.8n_{\rm H}\sqrt{2\pi m_{\rm H}kT_{\rm gas}}}{\gamma u_{\rm rad}\bar{\lambda}^{-2}}\right)^{1/2}\left(\frac{S_{\rm max}}{\rho}\right)^{1/4}(1+F_{\rm IR})^{1/2}\nonumber\\
&\simeq& 1.7 \left(\frac{\gamma_{-1}U}{n_{3}T_{1}^{1/2}}\right)^{-1/2}\left(\frac{\bar{\lambda}}{1.2\,\rm \mu m}\right) \hat{\rho}^{-1/4}S_{\rm max,7}^{1/4} (1+F_{\rm IR})^{1/2}\, \rm \mu m,
\label{eq:adisr_ana}
\end{eqnarray}
and the maximum size of grains that can still be disrupted by RAT-D is
\begin{eqnarray}
a_{\rm disr,max}&=&\frac{\gamma u_{\rm rad}\bar{\lambda}}{16n_{\rm H}\sqrt{2\pi m_{\rm H}kT_{\rm gas}}}\left(\frac{S_{\rm max}}{\rho}\right)^{-1/2}(1+F_{\rm IR})^{-1}\nonumber\\
&\simeq& 
0.03\left(\frac{\gamma_{-1} U}{n_{3}T_{1}^{1/2}}\right)\left(\frac{\bar{\lambda}}{1.2\,\rm \mu m}\right)\hat{\rho}^{1/2}S_{\rm max,7}^{-1/2}(1+F_{\rm IR})^{-1}\,\rm \mu m,\label{eq:adisr_up}
\end{eqnarray} 
which depend on the local gas properties, radiation field, and the grain tensile strength \citep{2020ApJ...891...38H,2021ApJ...908..218H}. 

As shown, the RAT-D efficiency strongly depends on the local conditions, i.e., more efficient for higher radiation strength (U) or $T_{\rm d}$, but less efficient for higher $n_{\rm H}$.

\subsubsection{Rotational desorption of ice mantles}
\begin{figure}
    \centering
    \includegraphics[width=0.7\textwidth]{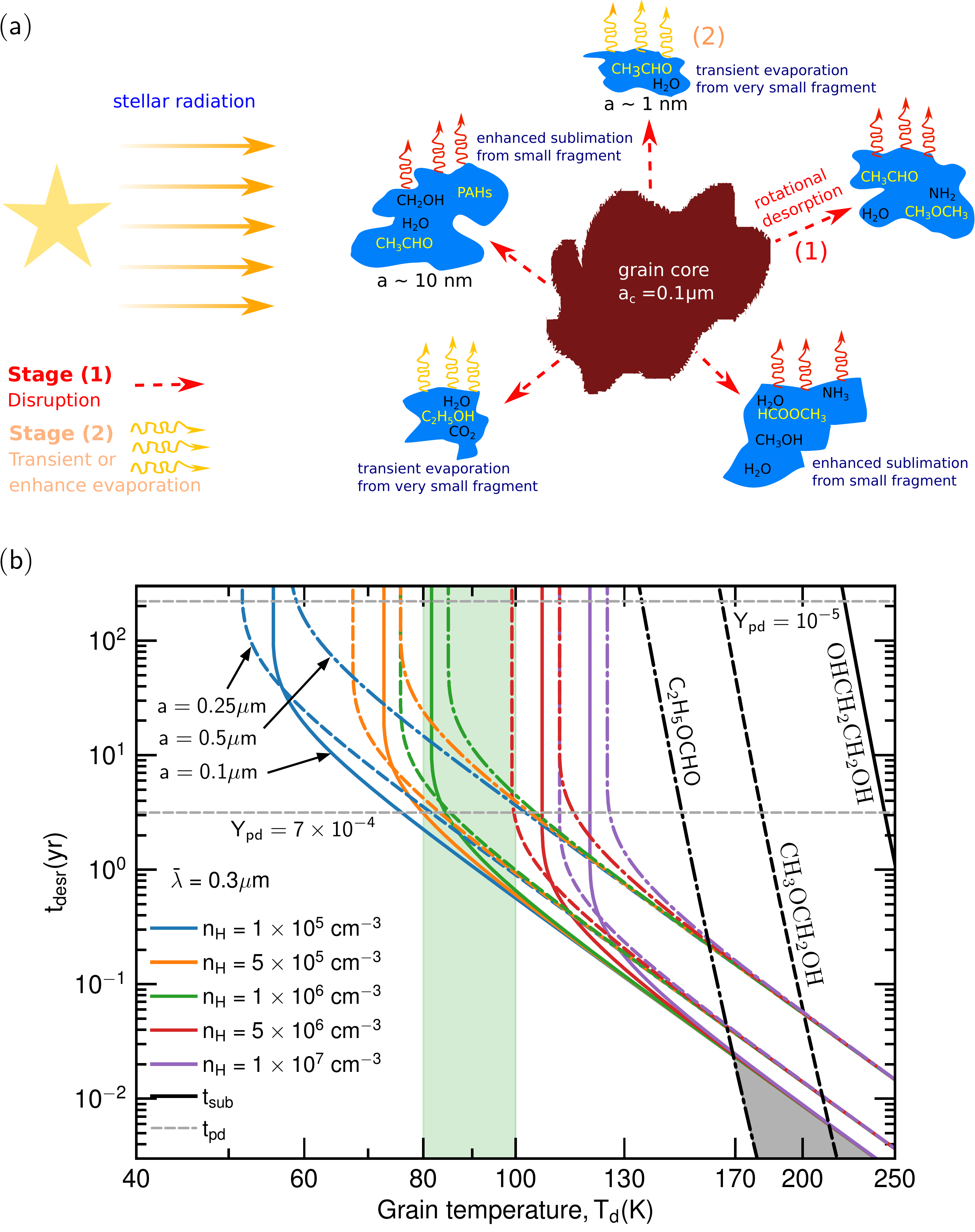}
    \caption{\textbf{(a)}:Illustration of the rotational desorption process (Figure adopted from \citealt{2020ApJ...891...38H}). There are two stages: (1) disruption of icy mantles into small fragments by the RAT-D mechanism, and (2) rapid evaporation due to thermal spikes for very small fragments or increased sublimation for larger fragments. The rotational desorption is thus able to desorb complex organic molecules (COMs) formed on the icy mantle. \textbf{(b)}: Desorption timescales of some COMs having very high binding energy in Orion BN/KL (Figure adopted from \citealt{2021ApJ...908..159T}). The rotational desorption and sublimation are shown by the colored and black lines, while the UV photodesorption is displayed by the gray dashed lines. The rotational desorption is effective for $T_{\rm d}\simeq 80-100\,\rm K$ at which these COMs are detected (cyan area). The gray area is where the sublimation is dominant.}
    \label{fig:rot_desorption}
\end{figure}

Detection of interstellar complex organic molecules (COMs) are extensively reported toward low-/high-mass protostars and protoplanetary disks (see e.g., \citealt[and references therein]{2014FaDi..168....9V}). As a popular scenario, COMs are believed to first form on the icy mantle of dust grains (e.g., \citealt{2008ApJ...682..283G}; \citealt{2016ApJ...830L...6J}), and subsequently desorbed to the gas phase during the warm and hot phase, where icy grain mantles can be heated to high temperatures $T_{\rm d}\sim 100–300\,\rm K$ (see e.g., \citealt{1987ApJ...315..621B}; \citealt{1988MNRAS.231..409B}; \citealt{2007A&A...465..913B}). However, icy mantles might not be survived under an intense radiation field before the critical temperature for thermal sublimation is reached (\citealt{2020ApJ...891...38H}). The centrifugal force can break the icy mantel from the bare core, then the evaporation and sublimation processes are enhanced owing to smaller size of the icy fragments, enabling COMs to release (see Figure \ref{fig:rot_desorption}(a)). This non-thermal process is called the rotational desorption, and found to be effective at lower dust temperatures than the classical thermal sublimation, which is able to explain the detection of COMs in cold regions (see \citealt{2020ApJ...891...38H,2021ApJ...908..159T} for further details). 

Figure \ref{fig:rot_desorption}(b) shows an evidence for rotation desorption of COMs from Orion BN/KL region. The timescales of rotational desorption (color lines) and thermal sublimation (dashed dotted black lines) for three COMs with very high binding energy, including ethyl formate ($\rm C_{2}H_{5}OCHO$), methoxymethanol ($\rm CH_{3}OCH_{2}OH$) and ethylene glycol ($\rm OHCH_{2}CH_{2}OH$) that are reported in \cite{2018A&A...620L...6T}, are shown for comparison. One can see that for given low dust temperatures constrained by SOFIA/FORCAST observations ($T_{\rm d}\simeq 80-100\,\rm K$; \citealt{2012ApJ...749L..23D}), rotational desorption is more effective than thermal sublimation in desorbing COMs from the grain surface. With regard to the UV photo-desorption (dashed horizontal lines), the rotational desorption could be as efficient as (or more effective than) the UV photo-desorption regarding to the photo-desorption yield uncertainty. However, UV photons are expected to be quickly attenuated by dust absorption, so that the volume in which photo-desorption is important is less extended than the rotational desorption which can work with optical-IR photons.

\begin{figure}
    \centering
    \includegraphics[width=0.85\textwidth]{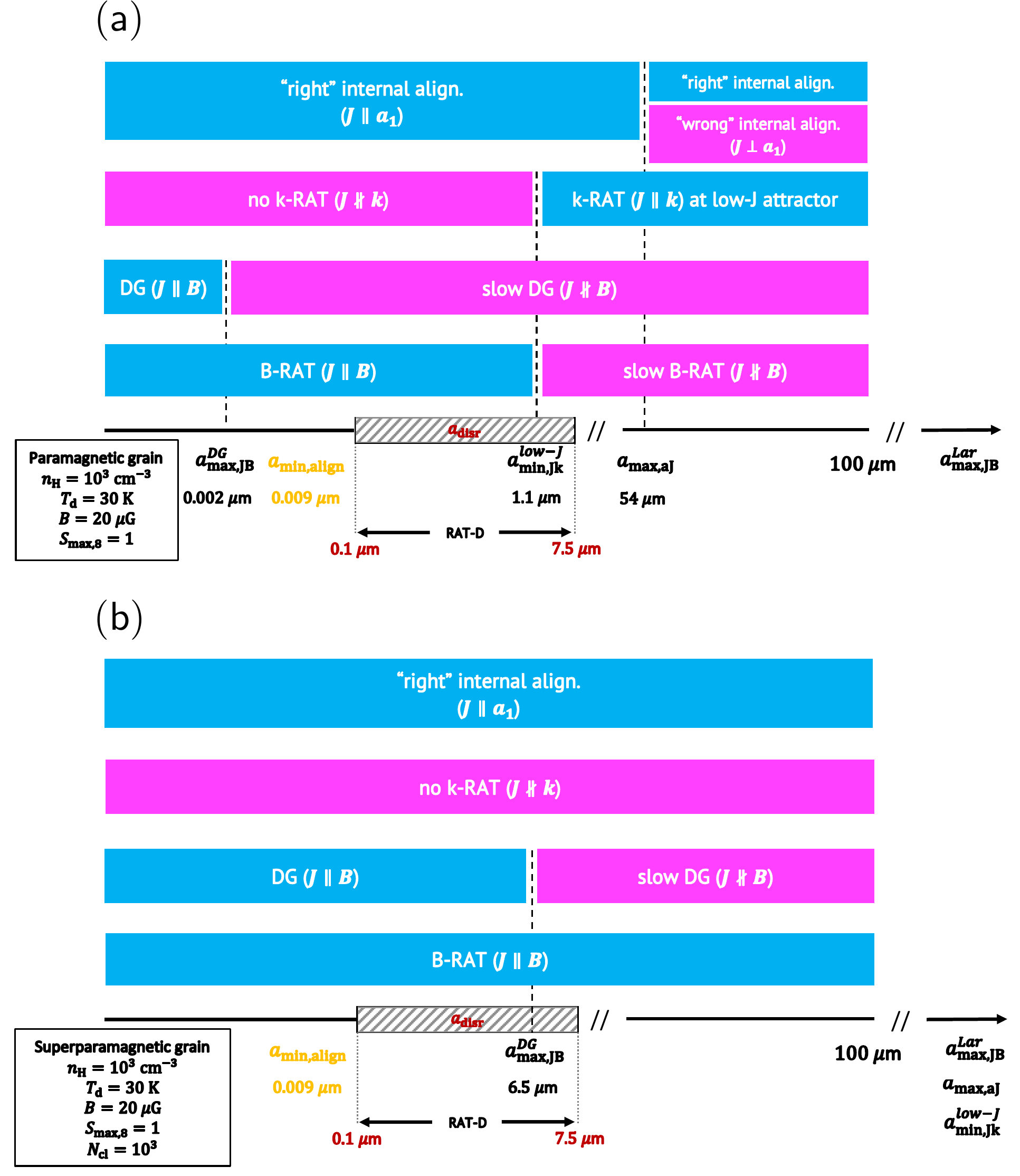}
    \caption{\textbf{(a)}: Alignment of paramagnetic grains for a typical molecular cloud in proximity to an ionization source with $\gamma=1$ and $\bar{\lambda}=0.5\,\rm \mu m$. The internal alignment of most grains is efficient (i.e., $\boldsymbol{a_{1}} \| \boldsymbol{J}$). For the external alignment, the paramagnetic relaxation (DG mechanism) is much less efficient than Larmor precession (B-RAT) to align $\boldsymbol{J}$ with $\boldsymbol{B}$. Grains with micro-size are aligned with magnetic fields. Larger grains are expected to align with radiation field (k-RAT). \textbf{(b)}: Similar but for superparamagnetic grains with $N_{\rm cl}=10^{3}$. Due to the enhancement of the grain susceptibility, (1) all grains are internally aligned with $\boldsymbol{J}$ perpendicular to $\boldsymbol{a_{1}}$, (2) DG mechanism is effective to larger grain size, (3) all grains are externally aligned with magnetic fields. Note that the B-RAT is much more effective than DG relaxation.}
    \label{fig:alignment_pm_spm}
\end{figure}
\subsection{Summary of grain alignment and disruption in molecular clouds}\label{sec:summary_align}
In this section, we discuss the alignment of magnetic grains in a typical molecular cloud, which is irradiated by an ionization source (e.g., OB stars). We adopted the following representative parameters: $n_{\rm H}=10^{3}\,\rm cm^{-3}$, $T_{\rm gas}=T_{\rm d}=30\,\rm K$, $U\simeq a^{6/15}_{-5}(T_{\rm d}/16.4\,\rm K)^{6}$, $\bar{\lambda}=0.5\,\mu$m, $\gamma=1.0$, and $B=20\,\rm \mu G$. The maximum grain size is limited to $100\,\mu$m.

Figure \ref{fig:alignment_pm_spm}(a) illustrates the relative importance of different alignment mechanisms. For the internal alignment, most grains are aligned with the angular momentum parallel to the short axis (i.e., $\boldsymbol{J} \parallel \boldsymbol{a_{1}}$). The paramagnetic relaxation is much less efficient than the Larmor precession. Grains can be aligned by RATs along B-fields (B-RAT) up to sizes of 1.1$\,\mu$m. Above that size, the radiative precession is faster than the Larmor precession such that grains are turned to align with the radiation direction (k-RAT). Hence, B-RAT is the most important alignment mechanism in the diffuse ISM and typical molecular clouds where dust grains could not grow to beyond $1\mu$m. Furthermore, if the grain's structure is not very compact (e.g., with a low tensile strength of $S_{\rm max}=10^{8}\,\rm erg\,cm^{-3}$), the RAT-D mechanism can disrupt large grains with size of 0.1-7.5$\,\mu$m. The rotational fragmentation enhances the abundance of smaller grains, which, in turn, can allow more grains to align with the magnetic field.

Similarly, Figure \ref{fig:alignment_pm_spm}(b) shows alignment for superparamagnetic grains with $N_{cl}=10^{3}$ and $\phi_{\rm sp}=0.01$. Due to the enhancement of susceptibility, all superparamagnetic grains are aligned with $\boldsymbol{J} \parallel \boldsymbol{a_{1}}$ for the internal alignment. For the external alignment, B-RAT alignment is dominant over k-RAT. Even the DG mechanism is more effective for larger grains up to 6.5$\,\mu$m compared to ordinary paramagnetic grains, yet the B-RAT is much more efficient than the superparamagnetic relaxation. Thus, superparamagnetic grains are precessed along the external magnetic fields. 

Therefore, grain alignment is able to probe the magnetic field in typical molecular clouds. However, as shown in this section, the gas damping, the B-field strength and the radiation strength are as key to characterize which alignments take place. Hence, all of alignment sizes are subjected to change with the local conditions, and the picture of grain alignment become very complicated in very dense regions, such as protostellar cores/disks (see Section \ref{sec:align_dense_gas}).

\section{Modelling thermal dust polarization within the RAT paradigm}\label{sec:modelling}
Polarized thermal dust emission depends on the grain properties (size, shape, and composition), grain alignment degree, and the magnetic field geometry (see e.g., \citealt{2009ApJ...696....1D}). Within the RAT paradigm, RATs can induce the simultaneous effects of RAT-A and RAT-D on dust grains. The RAT-A describes the alignment degree of grains of different sizes with the magnetic field, while RAT-D affects the grain size distribution. Therefore, an accurate physical model of polarized thermal dust emission requires the treatment of both RAT-A and RAT-D effects.

\subsection{Assumptions and basic equations}

\begin{figure}
    \centering
    \includegraphics[width=0.9\textwidth]{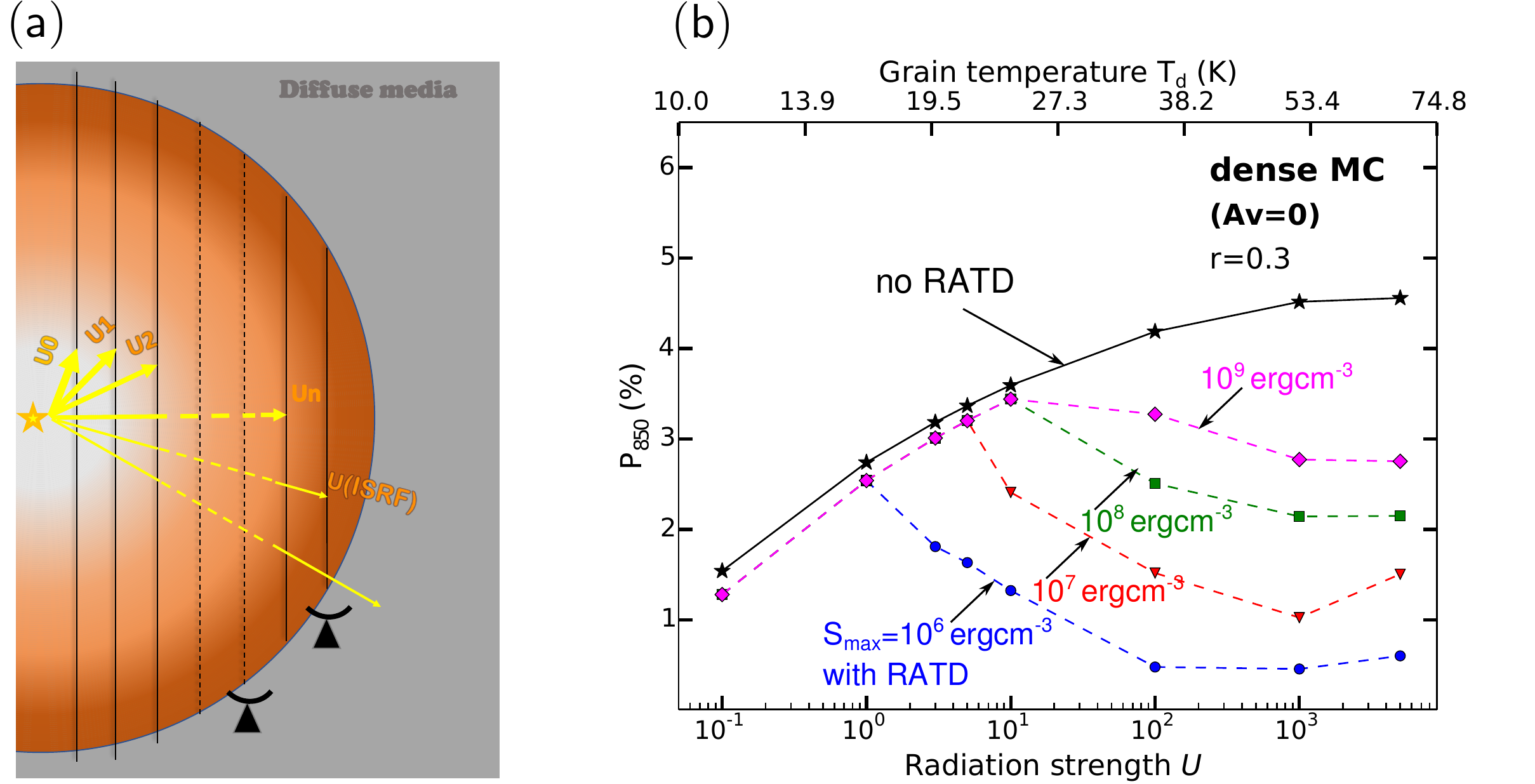}
    \caption{\textbf{(a)}: Schematic illustration of the model. A molecular cloud is irradiated by the central star, so that the radiation strength decreases from $U_{0}$ to $U_{\rm ISRF}$. The color gradient indicates the increase of the gas density from the central star. Magnetic fields are assumed to be uniform. \textbf{(b)}: An example of polarization degree vs. radiation strength (dust temperature) at $850\,\mu$m predicted by the numerical model (\citealt{2020ApJ...896...44L}). If we consider only the RAT-A effect, the polarization degree monotonically increases with increasing the radiation strength (dust temperature) as expected (black line). If we take into account the RAT-D mechanism, rotational disruption occurs when the radiation strength (dust temperature) is sufficiently high. Once this happens, the depletion of largest grains results in the decrease of the polarization degree (color lines). Figure taken from \cite{2020ApJ...896...44L} and \cite{2021ApJ...906..115T}.}
    \label{fig:model_sketch_prediction}
\end{figure}

To account for both the RAT-A and RAT-D effects on dust polarization, \cite{2020ApJ...896...44L} adopted a simple cloud model\footnote{The state-of-the-art version is publicly available at https://github.com/lengoctram/DustPOL} consisting of a homogeneous molecular cloud irradiated by a central radiation source, as illustrated in Figure \ref{fig:model_sketch_prediction}(a). The radiation strength ($U$) decreases with distance from the central star and varies along the line of sight (solid and dashed lines). For simplicity, all grains along each line of sight in the cloud are assumed to be illuminated by the constant radiation field of strength $U$.

As shown in Section \ref{sec:internal} (also Section \ref{sec:summary_align}), both Barnett and inelastic relaxation effects are very efficient for standard grains in the ISM and MCs, leading to the perfect internal alignment of dust grains (see Figures \ref{fig:alignment_pm_spm}). Therefore, grain alignment is mostly determined by external alignment by RATs. For the rotational damping, we consider only collisions with gas and re-emission of thermal dust as two main sources to damp the grain rotation against the spinup by RATs. The variation of B-fields along the line of sight is neglected by assuming the uniform magnetic field (the field tangling effect is discussed in Section \ref{sec:field_tangling}). The detailed description of this model was presented in \cite{2020ApJ...896...44L} and \cite{2021ApJ...906..115T}.

Dust grains are heated by absorption of the stellar radiation and cool down by re-emission in infrared radiation. Assuming a dust environment containing carbonaceous and silicate grains, the total dust emission intensity is given by
\begin{equation}
    \frac{I_{\rm em}(\lambda)}{N_{\rm H}} = \sum_{j=\rm sil,carb} \int^{a_{\rm max}}_{a_{\rm min}} Q_{\rm ext}(a,\lambda)\pi a^{2}
    \times\int dT B_{\lambda}(T_{\rm d})\frac{dP}{dT}\frac{1}{n_{\rm H}}\frac{dn_{j}}{da}da.~~~
\end{equation}
where $a_{\rm min}$ and $a_{\rm max}$ are the minimum and maximum grain sizes, $N_{\rm H}$ is the hydrogen column density, $B_{\lambda}(T_{\rm d})$ is the Planck function at dust temperature $T_{\rm d}$, $dP/dT$ is the grain temperature distribution function, $dn_{j}/da$ is the grain-size distribution of dust population $j$, and Q$_{\rm ext}$ is the extinction coefficient. The dust temperature distribution depends on the grain size, optical constant, and the radiation field ($U$), which is computed by the DustEM code (\citealt{2011A&A...525A.103C}, see e.g., Figure 8 in \citealt{2020ApJ...896...44L}).

If silicate and carbon dust populations are separated, then as paramagnetic grains, silicates can align with the ambient magnetic field, while carbonaceous grains cannot. Thus, the polarized emission intensity resulting from aligned grains is given by (\citealt{2016ApJ...831..159H})
\begin{equation}
    \frac{I_{\rm pol}(\lambda)}{N_{\rm H}}= \int^{a_{\rm max}}_{a_{\rm min}} f(a)Q_{\rm pol}\pi a^{2}
    \times\int dT B_{\lambda}(T_{\rm d})\frac{dP}{dT}\frac{1}{n_{\rm H}}\frac{dn_{sil}}{da}da, ~~~
\end{equation}
where $f(a)=1-\exp[-(a/2a_{\rm align})^{3}]$ is the alignment function, and Q$_{\rm pol}$ is the polarization efficiency coefficient.

If silicate and carbon grains are mixed together (e.g., \citealt{2013A&A...558A..62J,Draine.2021b1e}), which may be the case in dense clouds due to many cycles of photo-processing, coagulation, shattering, accretion, and erosion, carbon grains could be aligned with the ambient magnetic field and their thermal emission could be polarized. In the simplest case, assuming these grain populations have the same alignment parameters (i.e., $a_{\rm align}$ and $f(a)$), the total polarized intensity is
\begin{equation}
    \frac{I_{\rm pol}(\lambda)}{N_{\rm H}} = \sum_{j=\rm sil,carb} \int^{a_{\rm max}}_{a_{\rm min}} f(a)Q_{\rm pol}\pi a^{2}
    \times\int dT B_{\lambda}(T_{\rm d})\frac{dP}{dT}\frac{1}{n_{\rm H}}\frac{dn_{j}}{da}da.
\end{equation}

The extinction ($Q_{\rm ext}$) and polarization coefficients ($Q_{\rm pol}$) are computed by the DDSCAT model (\citealt{1994JOSAA..11.1491D, 2008JOSAA..25.2693D}; \citealt{2012OExpr..20.1247F}). The maximum grain size is constrained by the RAT-D mechanism, $a_{\rm max}\equiv a_{\rm disr}$.
 
The fractional polarization of the thermal dust emission is the ratio of the polarized intensity ($I_{\rm pol}$) to the total emission intensity ($I_{\rm em}$), which yields 
\begin{equation}\label{eq:pol_degree}
    P(\%) = 100\times \frac{I_{\rm pol}}{I_{\rm em}}.
\end{equation}

\subsection{Model parameters and numerical prediction}
Our model parameters include the gas properties (gas number density, $n_{\rm H}$) and gas temperature, $T_{\rm gas}$), the dust properties (size $a$, shape, internal structure described by the tensile strength $S_{\rm max}$, and size distribution power index $\beta$), and the ambient properties (radiation field strength $U$, mean wavelength $\bar{\lambda}$, and an anisotropy degree $\gamma$ of the radiation field). 

Radiation strength cannot be constrained from observations, thus, the equilibrium temperature of dust grains is used as a proxy of the radiation strength, as $U\simeq a^{6/15}_{-5}(T_{\rm d}/16.4\,\rm K)^{6}$. Here, the dust temperature is derived from fitting a modified black-body function to the continuum spectral energy distribution. The mean wavelength ($\bar{\lambda}$) of the stellar radiation field from a central source with $T_{\ast}$ is given as $\bar{\lambda}\simeq 0.53\,\rm cm\, K/T_{\ast}$ (see Eq. 8 in \citealt{2021ApJ...908..218H}). The internal structure of grains is determined via their tensile strength ($S_{\rm max}$). Unfortunately, this quantity is poorly constrained in astrophysical environments. However, compact grains in general are expected to have higher $S_{\rm max}$ than composite and fluffy grains. For instance, one expects $S_{\rm max} \sim 10^{9}-10^{10}\,\rm erg\,cm^{-3}$ for compact grains, while $S_{\rm max}\sim 10^{6}-10^{8}\,\rm erg\,cm^{-3}$ for composite grains (see e.g., \citealt{2020Galax...8...52H}). In the model, $S_{\rm max}$ can be set as a free parameter.

Figure \ref{fig:model_sketch_prediction}(b) shows the predicted dependence of the polarization degree on radiation strength (or dust temperature) at a specific wavelength of $850\,\mu$m for $n_{\rm H}=10^{4}\,\rm cm^{-3}$. In the case of only RAT-A effect (black line), the polarization degree increases as radiation strength (dust temperature) increases. In the case of both RAT-A and RAT-D effects, the polarization degree first increases and turns to decrease toward higher dust temperature. The second trend is caused by the rotational disruption which occurs above a critical temperature. For dust temperatures lower than the critical temperature, RATs are not enough to spin-up grains to the disruption limit $\Omega_{\rm disr}$, so that the RAT-D cannot occur. Thus, the results are the same as in the case of only RAT-A effect. For dust temperatures higher than the critical value, RATs become stronger to trigger the RAT-D effect, resulting in rotational disruption of large grains into small fragments. This leads to a drop in the polarization degree due to the depletion of large grains. This critical temperature and the level of the decline depend on the
internal structure of the grains controlled by $S_{\rm max}$.

\begin{figure}
    \centering
    \includegraphics[width=0.8\textwidth]{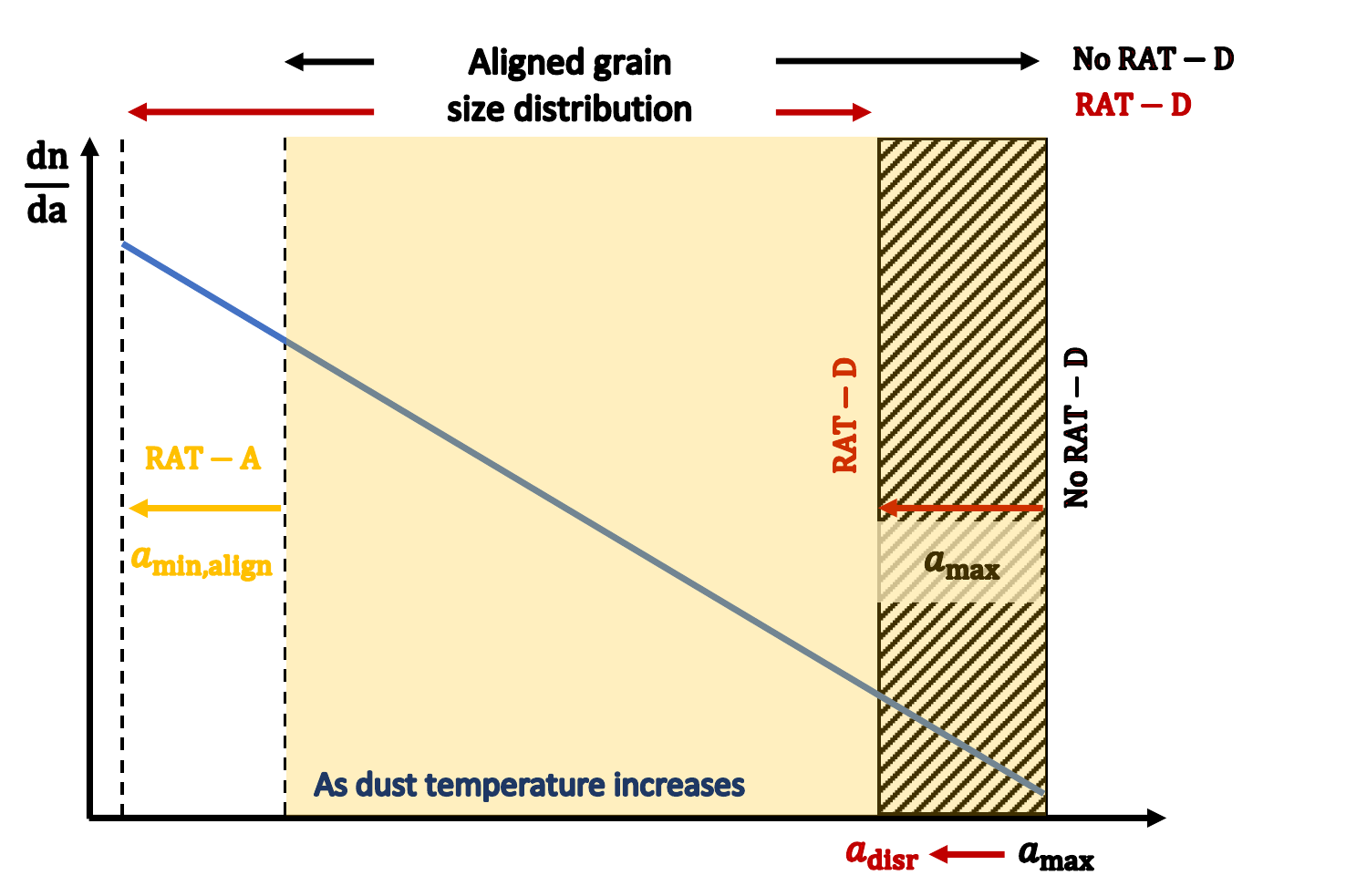}
    \caption{Understanding of thermal dust polarization. For a fixed maximum grain size ($a_{\rm max}$), higher radiation strength/dust temperature can bring smaller grains to align (smaller size of $a_{\rm min,align}$) so that the aligned grain size distribution becomes broader. This broad size distribution increases the degree of thermal dust polarization with radiation strength/dust temperature. If the maximum grain size is determined by the RAT-D mechanism and grains are composite ($a_{\rm max}\rightarrow a_{\rm disr}$), higher radiation/dust temperature results in smaller $a_{\rm disr}$. The consequent narrower aligned grain size distribution drops the polarization degree.}
    \label{fig:model_understanding}
\end{figure}

Figure \ref{fig:model_understanding} shows our understanding of the trend of polarization degree versus radiation strength/dust temperature shown in Figure \ref{fig:model_sketch_prediction}(b). For a given gas density, the polarization degree is mainly determined by the grain size distribution. If the only RAT-A effect is considered, the maximum grain size is fixed. The minimum size that is aligned by RATs is smaller for higher radiation strength (more small grains are able to be aligned). Hence, the size distribution of aligned grains becomes broader, which deduces the increment of the polarization degree. If the rotational disruption is taken into account along with the RAT-A effect, the maximum grain size could vary. In this case, if the radiation is insufficient strong the polarization degree is observed to increase with radiation strength as previous case. Once the radiation strength is sufficient to trigger disruption, the maximum grain size reduces from $a_{\rm max}$ to $a_{\rm disr}$. The narrower size distribution of aligned grains causes the polarization degree to drop.

As we mentioned, radiation strength is not an observable quantity, thus dust temperature could be used as a proxy. Therefore, the relation between polarization degree and dust temperature ($p-T_{\rm d}$) becomes a unique tool to study the joint effect of grain alignment and disruption by RATs.

\section{Observational Study of the RAT effects in star-forming regions}\label{sec:obs}
In this section, we review our observational results of thermal dust polarization observed toward star-forming molecular clouds, including $\rho$ Ophiuchus-A, 30 Doradus, M 17, and Orion BN/KL. These star-forming MCs containing intense radiation sources are the ideal targets to test the RAT-A and RAT-D effects.

\begin{figure}
    \centering
    \includegraphics[width=0.75\textwidth]{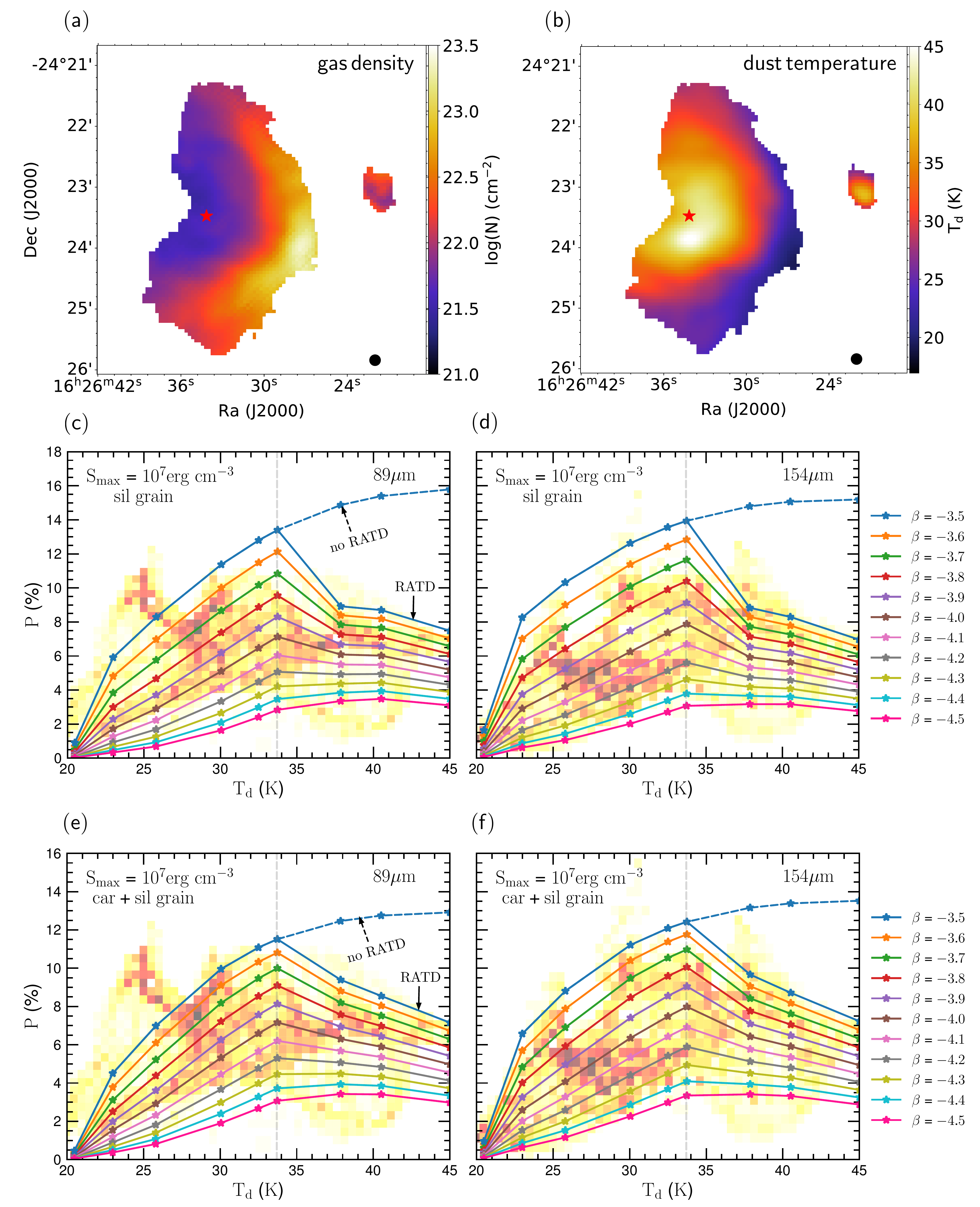}
    \caption{\textbf{Top}: Gas column density (panel (a)) and dust temperature (panel (b)) constructed from Herschel data (Figure adopted from \citealt{2019ApJ...882..113S} and \citealt{2021ApJ...906..115T}). A star symbol locates the radiation source Oph-S1. Close to this source, dust is hotter while gas is more diffuse. \textbf{Middle}: 2D-histogram of polarization degree observed with SOFIA/HAWC+ at 89$\,\mu$m (panel (c)) and 154$\,\mu$m (panel (d)) and dust temperature (background). Dashed line is the prediction by only the RAT-A effect, which could not explain the decrease of $p$ for $T_{\rm d}>34\,\rm K$. The solid lines are the predictions made by both the RAT-A and RAT-D effects for only silicate grains with the grain-size distribution of $dn/da\sim a^{\beta}$. \textbf{Bottom}: Similar to the middle panel but for and a mixture of silicate-carbonaceous grains at 89$\,\mu$m (panel(e)) and 154$\,\mu$m (panel(f)). The RAT paradigm (RAT-A and RAT-D) successfully reproduces both the increase and decrease part of the data. Figures adopted from \cite{2021ApJ...906..115T}.}
    \label{fig:OphA-fit2obs}
\end{figure}
\subsection{$\rho$ Ophiuchus-A}
$\rho$ Ophiuchus-A ($\rho$ Oph-A) is a molecular cloud in one of the closest dark cloud complex and star-forming region $\rho$ Ophiuchi. The distance to this complex is reported to be $\sim$ 120--160 pc \citep{1981A&A....99..346C, 1989A&A...216...44D, 2017ApJ...834..141O}. This cloud is irradiated by a high-mass star, namely Oph-S1, which is a young B3-type star (\citealt{1977AJ.....82..198V}; \citealt{2003PASJ...55..981H}).

Figure \ref{fig:OphA-fit2obs} shows the observational data of $\rho$ Oph-A. The top panels show the maps of gas column density (panel (a)) and dust temperature (panel (b)) constructed from {\it Herschel}. Along the radial direction toward Oph-S1 (marked by the white star), the dust is hotter while the gas is more diffuse.
The middle and bottom panels show the $p-T_{\rm d}$ relations. Background shows the observational data at 89 (left) and 154$\,\mu$m (right) observed by SOFIA/HAWC+. The polarization degree tends to first increase and then decrease as the dust temperature increases. One can see that the latter decrement occurs in vicinity of Oph-S1 where the gas density is lower. Therefore, the drop of $p$ toward higher $T_{\rm d}$ and lower $n_{\rm H}$ is opposite to the fundamental prediction of the RAT-A effect.

The lines in Figure \ref{fig:OphA-fit2obs} show the results from our numerical modelling overplotted with the observational data at two wavelengths, assuming only silicate (panels (c,d)) or a combination of silicate and carbonaceous grains (panels (e,f)) with different slopes ($\beta$) of the size distribution. We considered two cases: only the RAT-A effect (no RAT-D, dashed line) and both RAT-A and RAT-D effects (solid lines).
\begin{itemize}
    \item For the former case, the polarization degree monotonically increases with the dust temperature. This rising $p-T_{\rm d}$ trend is caused by the increase in the alignment efficiency (i.e., decrease in the alignment size $a_{\rm align}$) toward the Oph-S1 due to stronger radiation intensity and lower gas density. Obviously, the RAT-A effect is able to reproduce the first increasing part of the observational data, but not the second decreasing part.
    \item For the latter case, the polarization degree is similarly expected to increases for $T_{\rm d}\simeq 34\,\rm K$. At higher dust temperatures, the RAT-D happens for grains larger than $a_{\rm disr}$ owing to the higher radiation intensity and the lower gas density, which reduces the polarization degree. The joint effect of RAT-A and RAT-D mechanisms is therefore able to reproduce the entire observational data.
\end{itemize}
 
\subsection{30 Doradus}
30 Doradus (30 Dor), the Tarantula Nebula, is an \textsc{Hii} region located in the Large Magellanic Cloud (LMC) at the distance of
$\sim 50\,\rm kpc$ from the Sun (e.g., \citealt{2011ApJ...739...27D}). It is powered by the massive star cluster R$\,$136 that has a
bolometric luminosity of $7.8\times 10^{7}L_{\odot}$ (see \citealt[and references therein]{2011ApJ...731...91L}). Since the LMC is a low-metallicity
galaxy ($Z\simeq 0.5Z_{\odot}$, see e.g., \citealt{1982ApJ...252..461D}; \citealt{2008ApJ...679..310G}) that has a low dust-to-gas ratio (e.g.,
$\sim 2–5\times 10^{-3}$; see \citealt{2014ApJ...797...86R}), the radiative feedback could be effective deeper inside the surrounding molecular cloud. Thus, the 30 Dor cloud offers a valuable environment to test the physics of grain alignment and disruption by RATs.
\begin{figure}
    \centering
    \includegraphics[width=0.9\textwidth]{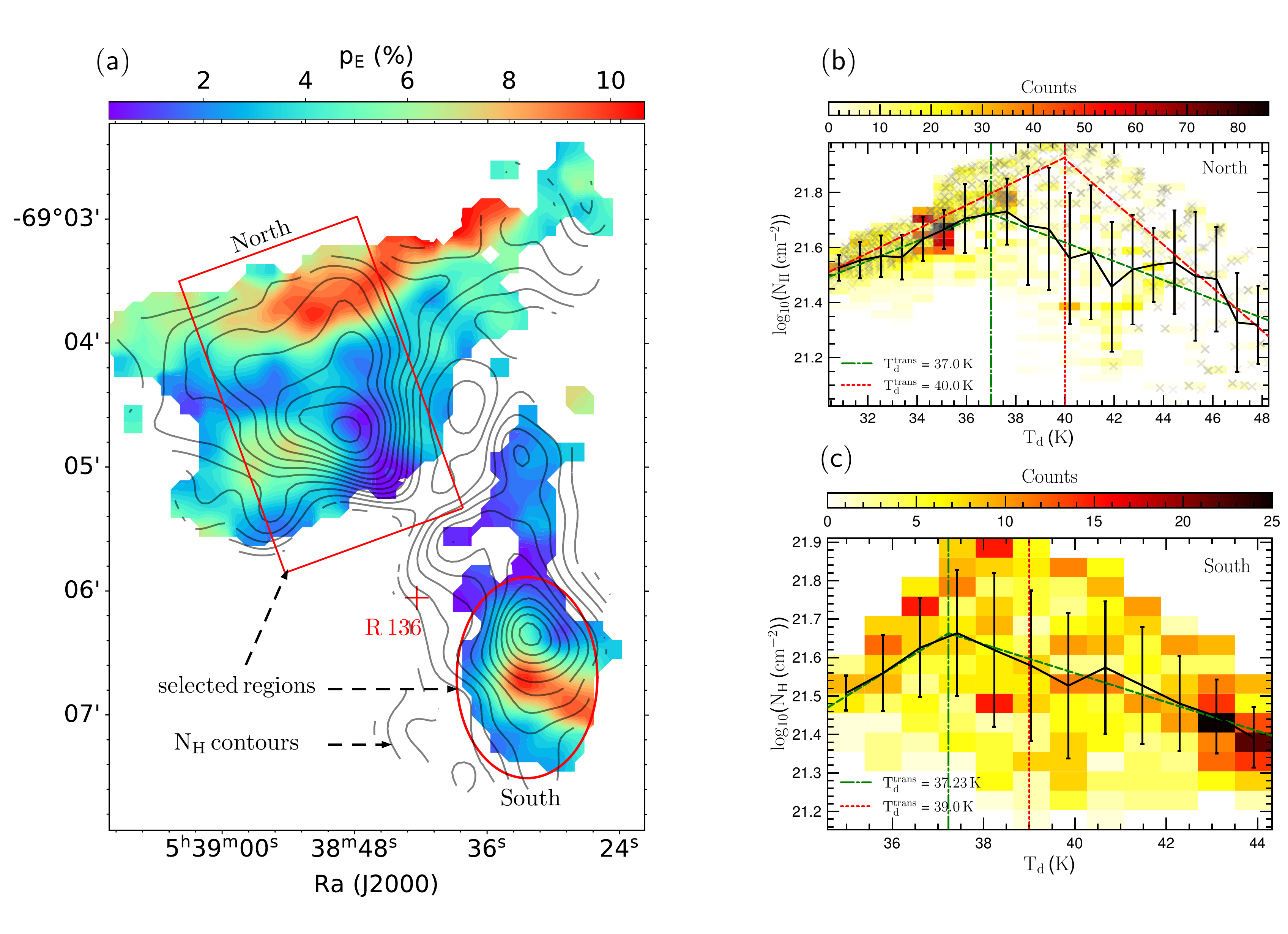}
    \caption{\textbf{(a)}: SOFIA/HAWC+ observations of 30 Dor at 214$\,\mu$m. Contours are the map of gas column density. The stellar cluster R$\,$136 is located by a red cross. We selected two sub-regions, namely North and South, to further analysis. \textbf{(b,c)}: 2D-histogram of gas column density vs. dust temperature in two selected sub-regions: North and South. The black line shows the weighted mean in each bin associated with $1-\sigma$ uncertainty. The blue line is a double power-law fitting. In the North, we fit to an upper envelope of the distribution by a dashed red line. In the South, the plot is more scatter. However, these quantities are positively correlated upto $T_{\rm d}\simeq 37\,\rm K$, and negatively correlated to higher $T_{\rm d}$. Figure adopted from \citealt{2021ApJ...923..130T}.}
    \label{fig:30Dor_hawc+}
\end{figure}

Figure \ref{fig:30Dor_hawc+}(a) shows the polarization degree map observed with SOFIA/HAWC+ at 214$\,\mu$m. The location of ionization source R$\,$136 is denoted by a red cross. For comparison, the contours show the map of gas column density. For further analysis, we selected two sub-regions, namely North and South as highlighted respectively by a red box and a red ellipse. Figure \ref{fig:30Dor_hawc+}(b,c) show the relation of the gas column density to the dust temperature in these two sub-regions, respectively. In the North, the gas density unambiguously increases and then decreases with increasing dust temperature. The transition is at $T_{\rm d}\sim 37-40\,\rm K$. In the South, the relation is less clear. Yet, it is likely to behave similarly as in the North, in which the gas density positively correlates with the dust temperature upto $T_{\rm d}\sim 37-39\,\rm K$, and negatively correlates with higher $T_{\rm d}$.
\begin{figure}
    \centering
    \includegraphics[width=0.8\textwidth]{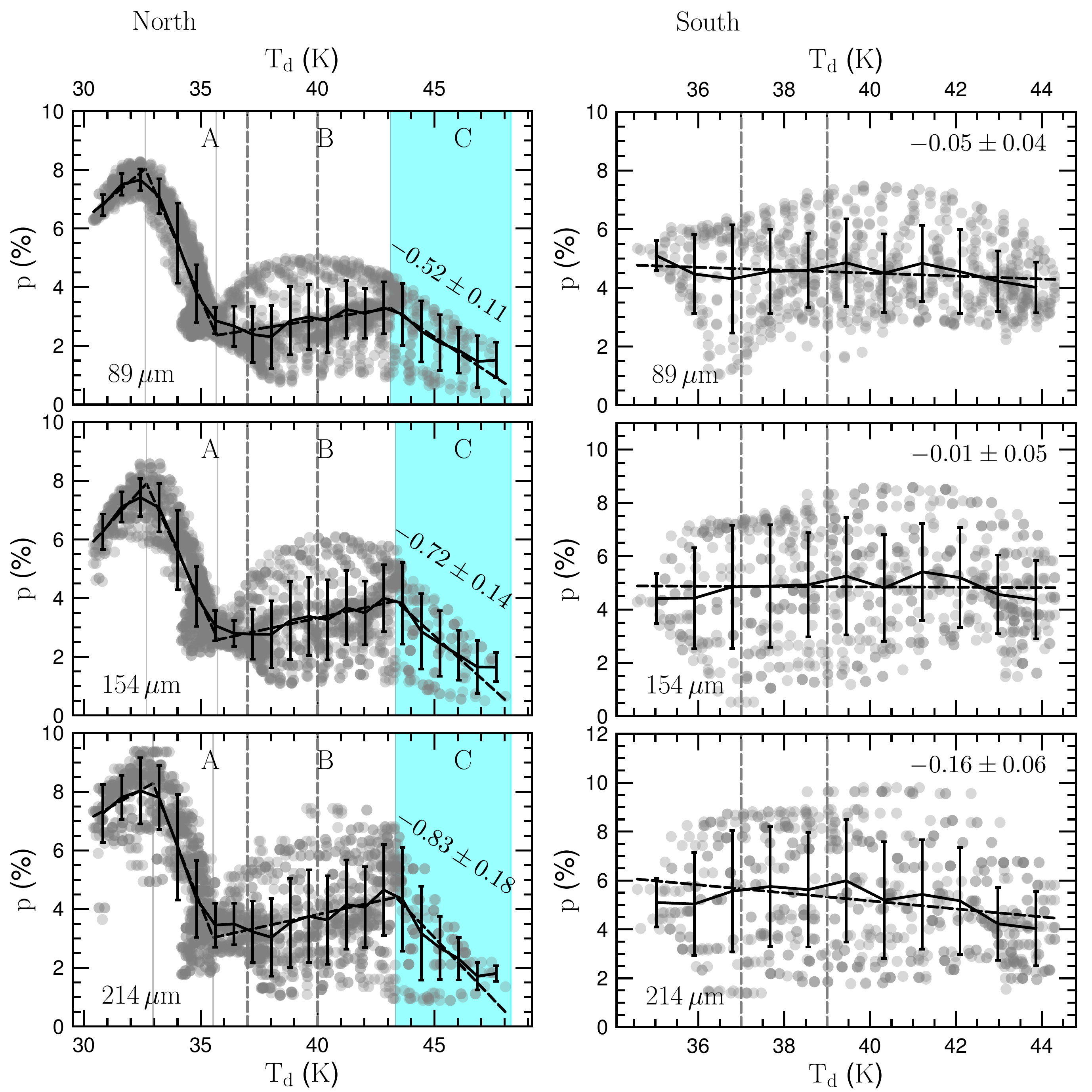}
    \caption{Relations between polarization degree and dust temperature in three bands (from top to bottom) in North (left panel) and in South (right panel) regions highlighted in the left panel of Figure \ref{fig:30Dor_hawc+}. The error bars connected by a line shows the weighted-mean in each bin. The black dashed line shows the piecewise line fit to the data. In the North, a complex $p-T_{\rm d}$ relation is found, but $p$ decreases with higher $T_{\rm d}$. In the South, a weak $p-T_{\rm d}$ relation is seen. Figure adopted from \citealt{2021ApJ...923..130T}.}
    \label{fig:30Dor_pTd}
\end{figure}

Figure \ref{fig:30Dor_pTd} shows the $p-T_{\rm d}$ relations in the North (left panel) and in the South (right panel) for three SOFIA/HAWC+ wavelengths (from top to bottom).
\begin{itemize}
    \item[1-] In the North, there is a complex relation for all wavelength. Note that $T_{\rm d}$ increases toward R$\,$136.
    \begin{itemize}
        \item For $T_{\rm d}\leq 37–40\,\rm K$ (region A), both dust temperature and gas column density increase. The polarization degree first rapidly drops, then rises, and finally slowly changes with increasing dust temperature. This trend could be explained by the RAT-A effect. Increasing gas density causes increased rotational damping by gas collisions, which is sufficiently efficient for larger $a_{\rm align}$. As we discussed in Section \ref{sec:modelling}, a higher value of $a_{\rm align}$ causes $p$ to decrease. At some point in the cloud where $T_{\rm d}$ is sufficient high, the variation of $a_{\rm align}$ might be small because of opposite effects of $T_{\rm d}$ and $N_{\rm H}$. That makes $p$ to slowly change.
        
        \item For $T_{\rm d}>37–40\,\rm K$, the gas column density decreases, the polarization degree first increases slightly (regions B) and then drops again (region C). The negative slopes are steeper for longer wavelengths, namely $-0.52$, $-0.72$ and $-0.83$ at 89, 154 and 214$\,\mu$m, respectively. As gas damping become less effective due to lower $N_{\rm H}$ and higher $T_{\rm d}$, the RAT-A effect could explain the increase in $p$ in region B (i.e., smaller $a_{\rm align}$), but fails to explain the drop of $p$ in region C. The latter is evidence for the RAT-D mechanism at work in proximity to radiation source where $N_{\rm H}$ continues to be small while $T_{\rm d}$ is highest. As the largest grains get disrupted, $p$ decreases faster at longer wavelength. Thus, the observed drops of $p$ in region C are expected by RAT-A and RAT-D.
    \end{itemize}
    
    \item[2-] In the South, the $p-T_{\rm d}$ relation has a high degree of scatter and is ambiguous. There might be three reasons for the slow variation of $p$ with $T_{\rm d}$. First, the saturation of RAT-D, which occurs at very high temperatures (i.e., $T_{\rm d}$ is much larger than the disruption threshold temperature; see Figure 14 in \citealt{2020ApJ...896...44L}). Small grains resulting from the disruption do not experience RAT-D due to lower RATs. Second, when dust grains are very hot, IR damping is so strong and dominates the gas damping, resulting in the slow variation of $a_{\rm disr}$ with $T_{\rm d}$. Third, effective gas damping due to the high gas density in the selected region (see Figure \ref{fig:30Dor_hawc+}). 

\end{itemize}

\subsection{M17}
M17, also known as the Omega Nebula or the Horseshoe Nebula is prominent \textsc{Hii} region with and associated molecular cloud and star-forming region (\citealt{Povich_2009}; \citealt{Lim_2020}). It is located in the constellation of Sagittarius at a distance of
1.98 kpc (\citealt{Xu_2011}). M17 hosts the stellar cluster NGC 6618 and its relative closeness makes it an excellent target to test the RAT paradigm.
\begin{figure}
    \centering
    \includegraphics[width=0.9\textwidth]{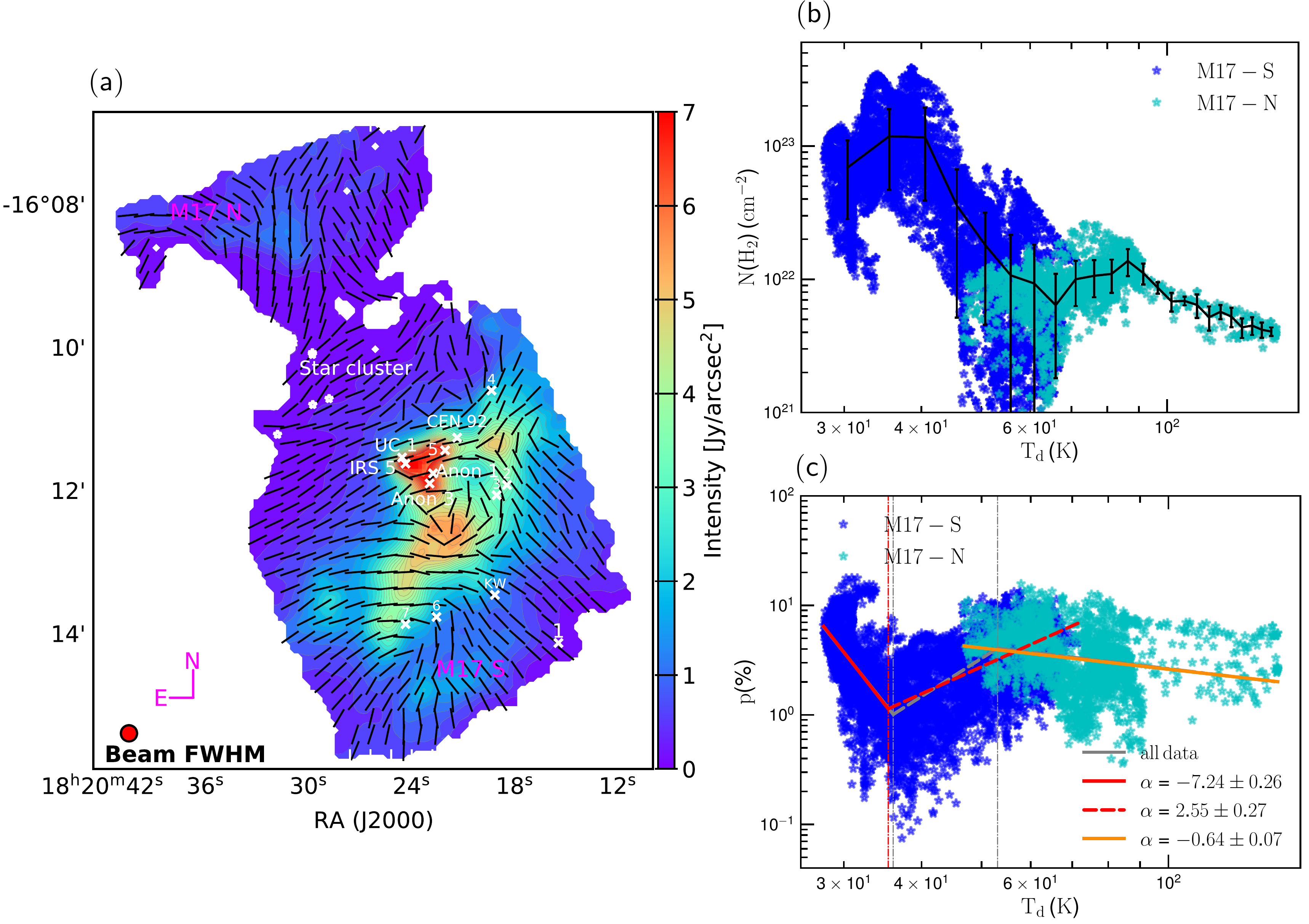}
    \caption{\textbf{(a)}: SOFIA/HAWC+ observations of M17. The color background shows the total continuum intensity (Stokes-I). The black vectors represent the magnetic field morphology. We devided M17 into two sub-regions: M17-N and M17-S. In the M17-S region, IR sources are denoted by the white crosses and numbered. Star cluster NGC 6618 is represented by star symbols. \textbf{(b)}: The relation of gas column density vs. dust temperature in the two sub-regions. \textbf{(c)}: The relation of polarization degree vs. dust temperature. Error bars connected by a line shows the weighted-mean in each bin and illustrates that the lower density gas has a higher dust temperature. Color lines are the fitting to all data point (the grey line), to only M17-S (the red line), and to only M17-N (the orange line). The polarization increases with increasing dust temperature in M17-S, while it drops to higher $T_{\rm d}$ and lower $N(\rm H_{2})$ in M17-N. Figure adopted from \cite{2022ApJ...929...27H}.}
    \label{fig:M17}
\end{figure}

Figure \ref{fig:M17}(a) shows the SOFIA/HAWC+ observations of M17. Generally, the stellar cluster NGC 6618 (the most massive stars in this cluster are indicated by the white star symbols) are responsible for ionizing and creating the \textsc{Hii} region. We divided M17 into two sub-regions, namely M17-N and M17-S. M17-S contains several compact sources as marked by white crosses. Figure \ref{fig:M17}(b,c) respectively show the relation between gas column density and dust temperature, and the relation of polarization degree with dust temperature. Lower density gas is associated with higher dust temperature.

\begin{itemize}
    \item In M17-S, where the density is relatively high, and the dust temperature is relatively low, the observed polarization degree increases with increasing dust temperature. In this condition of high gas density and low temperatures, gas damping is more efficient so that RAT-D cannot occur. Therefore, the increase of $p-Td$ is consistent with the prediction of the RAT-A effect. 
    
    \item In M17-N, where the gas density is rather low and the dust dust temperature is relatively high ($T_{\rm d}\geq 60\,\rm K$), the observed polarization tends to decrease with $T_{d}$. In this condition of lower densities and high temperatures, RAT-D can easily occur due to weak gas damping and strong RATs. As a result, the decrease in $p$ with increasing $T_{\rm d}$ is consistent with the joint effect of RAT-A and RAT-D.
\end{itemize}

\subsection{OMC-1 or Orion BN/KL}
\begin{figure}
    \centering
    \includegraphics[width=0.9\textwidth]{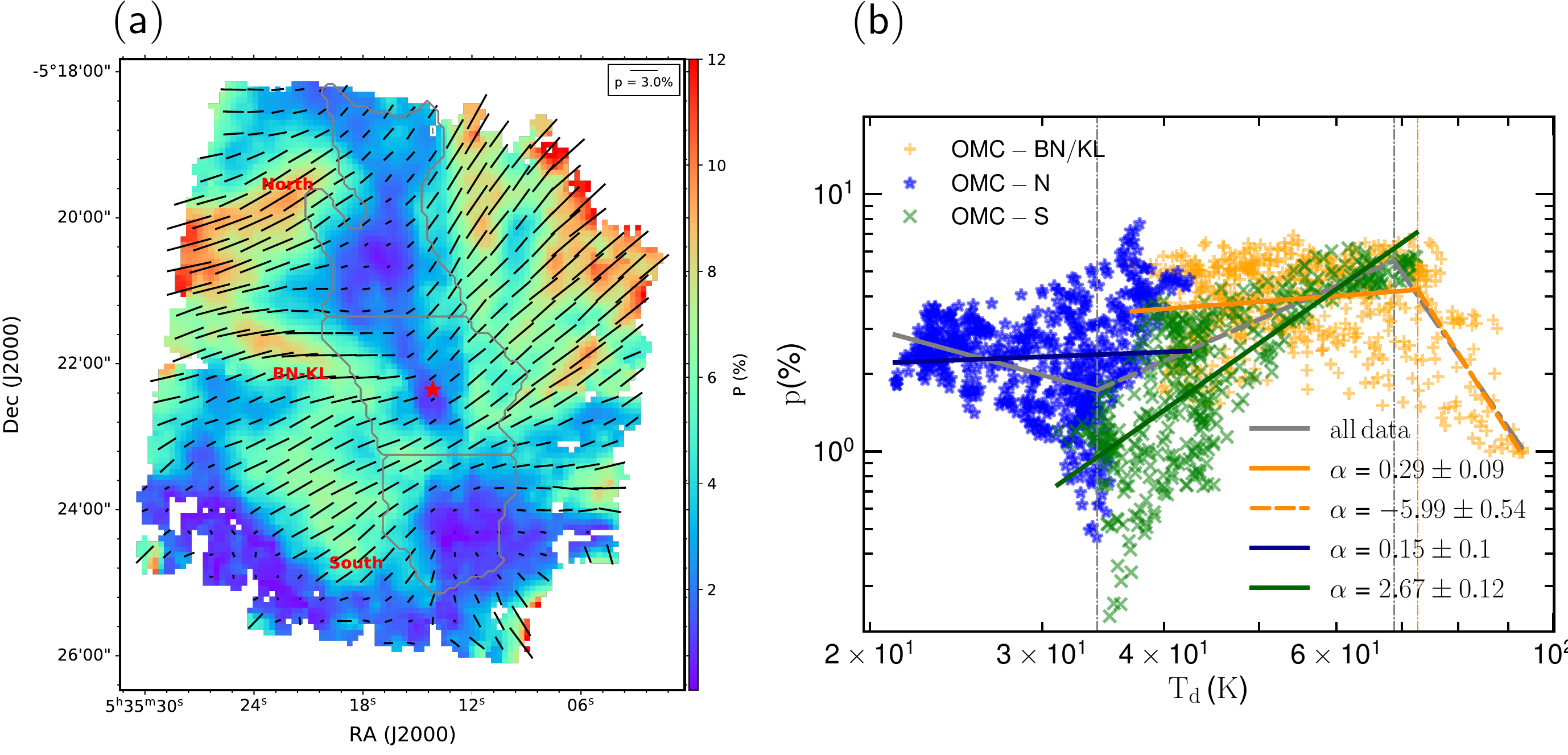}
    \caption{\textbf{(a)}: SOFIA/HAWC+ observations of OMC-1 at 214$\,\mu$m. The background color is polarization degree. The black vectors illustrate the B-field morphology inferred from thermal dust polarization (Ngoc et al. in preparation). Three sub-regions, namely BN-KL, North and South are selected for further analysis. \textbf{(b)}: The relation of polarization degree at 214$\,\mu$m to dust temperature at three sub-regions in OMC-1 (Ngoc et al. in preparation). In OMC-N and OMC-S, $p$ is seen to monotonically increases with $T_{\rm d}$. The increment in the first sub-region is faster than the latter one. It differs from OMC-BN/KL, for which $p$ slightly increases before deeply decreasing to higher $T_{\rm d}$.}
    \label{fig:Orion_pTd}
\end{figure}

At a distance of $\simeq 400\,$pc (\citealt{2007A&A...474..515M}; \citealt{kounkel2017oriondistance}), the Orion nebula is the closest and the most extensive studied region of massive star-forming region. Due to its proximity, the arcsecond resolution observations of polarized thermal dust that have been carried out sample scales of a few thousand AU (\citealt{2017ApJ...846..122P}; \citealt{2019ApJ...872..187C};\citealt{2022EPJWC.25700002A}). These observations have provided a wealth of information of dust physics in the different physical environments of the Orion region. Here we focus on only OMC-1 region. This region can be divided into: (1) the Becklin-Neugebauer-Kleinmann-Low Nebula (BN/KL; \citealt{becklinneugebauern1967,kleinmannlow1967}) which hosts on-going star formation; (2) the Southern clump (OMC-S; \citealt{batria1983orions}) that is quiescent. These sub-regions are labelled in Figure \ref{fig:Orion_pTd}(a). We concentrated on the variation of the polarization degree along the main filament in Orion, in which we selected one more sub-region (OMC-N) north of BN/KL.

Figure \ref{fig:Orion_pTd}(b) shows the variation of the polarization degree with the dust temperature for these three sub-regions. In two quiescent OMC-N and OMC-S, $p$ increases with increasing $T_{\rm d}$, which is predicted by the RAT-A effect. Since $T_{\rm d}$ is higher in OMC-S, the grain alignment is more efficient resulting in more significant increment in this region compared to OMC-N, which is understandable in the RAT-A framework. However, the $p-T_{\rm d}$ relation in OMC-BN/KL is different from other regions. The polarization degree slightly increases and then decreases for $T_{\rm d}\geq 70\,\rm K$. At such high $T_{\rm d}$, the rotational disruption is effective, leading to the decline of $p$. The RAT paradigm is therefore able to reproduce the $p-T_{\rm d}$ relations in these three regions of OMC-1.

\subsection{Can the anti-correlation of $p$ vs. $T_{\rm d}$ arise from B-field tangling?}\label{sec:field_tangling}
We showed that the anti-correlation of $p$ vs. $T_{\rm d}$ is produced by RAT-D mechanism, assuming uniform magnetic fields. This disregard of the magnetic field tangling on variation of $p$ could lead to biased conclusions. To check whether or not the field tangling could cause the decrease of $p$, we analyze the variation of the polarization angle dispersion function $S$ as a function of $T_{\rm d}$. The biased dispersion at position $\boldsymbol{r}$ on the sight-line is given as
\begin{equation}
    S_{\rm biased}(\boldsymbol{r}) = \sqrt{\frac{1}{N}\sum^{N}_{i=1}[\psi(\boldsymbol{r}+\boldsymbol{\delta}_{i})-\psi(\boldsymbol{r})]^{2}}
\end{equation}
where $\psi(\boldsymbol{r})$ and $\psi(\boldsymbol{r}+\boldsymbol{\delta}$) are the polarization angles at position $\boldsymbol{r}$ and $\boldsymbol{r}+\boldsymbol{\delta}$, respectively. In practice, for a given position $\boldsymbol{r}$ we select all data points ($N$) within a circular aperture centered at this position with a diameter of two beam sizes.

Then, the unbiased dispersion function is
\begin{equation}
    S(\boldsymbol{r}) = \sqrt{S^{2}_{\rm biased}(\boldsymbol{r})-\sigma^{2}_{S}(\boldsymbol{r})}
\end{equation}
whose an associated error is (see Eq. 8 in \citealt{2020A&A...641A..12P})
\begin{eqnarray}
    \sigma_{S}^{2}(\boldsymbol{r}) &=& \frac{1}{N^{2}S^{2}}\sum_{i=1}^{N}\sigma^{2}_{\psi}(\boldsymbol{r}+\boldsymbol{\delta_{i}})[\psi(\boldsymbol{r}+\boldsymbol{\delta}_{i})-\psi(\boldsymbol{r})]^{2} \nonumber \\
    &&+\frac{\sigma^{2}_{\psi}(\boldsymbol{r})}{N^{2}S^{2}}\left[\sum_{i=1}^{N}\psi(\boldsymbol{r}+\boldsymbol{\delta}_{i}) - \psi(\boldsymbol{r})\right]^{2}
\end{eqnarray}
with $\sigma(\boldsymbol{r})$ and $\sigma(\boldsymbol{r}+\boldsymbol{\delta})$ the errors of polarization angle at the position $\boldsymbol{r}$ and $\boldsymbol{r}+\boldsymbol{\delta}$.
\begin{figure}
    \centering
	\includegraphics[width=1\textwidth]{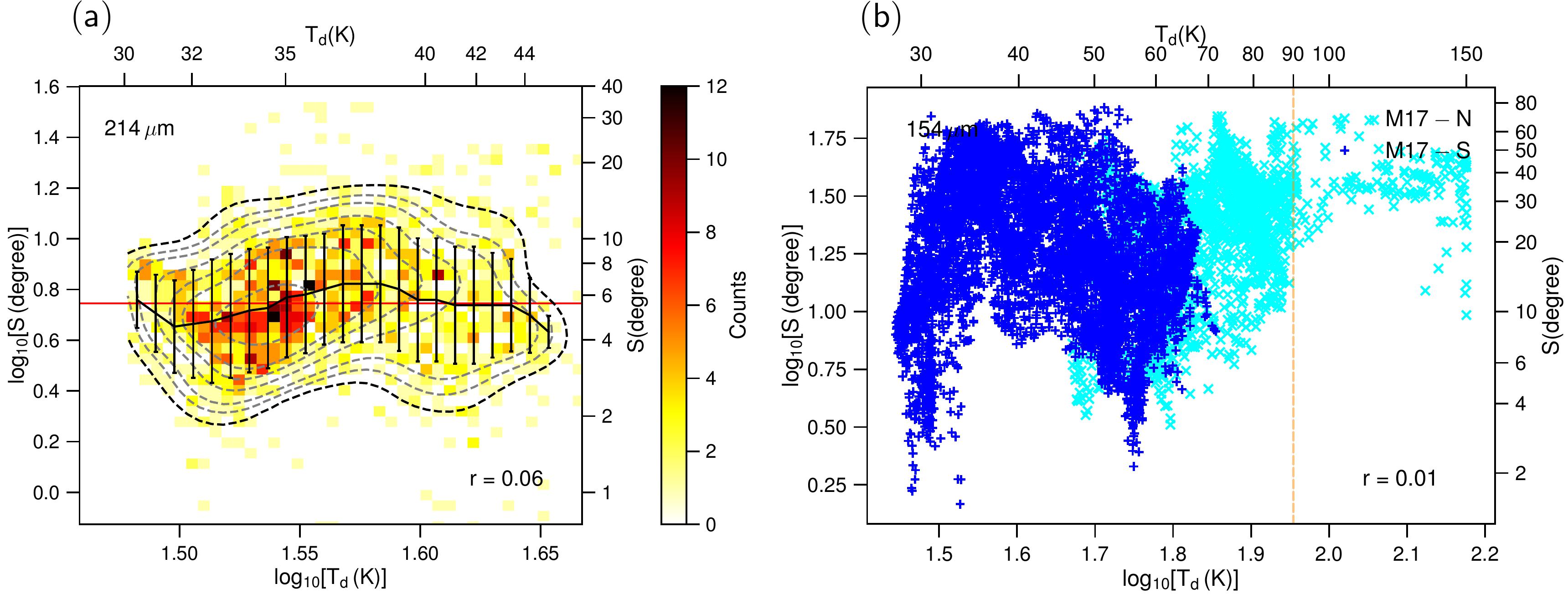}
    \caption{\textbf{(a)}: 2D-histogram of $S$ and $T_{\rm d}$ in 30 Dor at 214$\,\mu$m (Figure adopted from Tram et al. in prep.). They have a loose relation with a correlation coefficient of $r=0.06$. Thus, the variation of magnetic field could not be a main reason for a decrease in $p$ with $T_{\rm d}>43\,\rm K$ as shown in Figure \ref{fig:30Dor_pTd}. \textbf{(b)}: Scatter plot of $S$ vs. $T_{\rm d}$ for M17 at 154$\,\mu$m (Figure adopted from \citealt{2022ApJ...929...27H}) with a correlation coefficient of $r=0.01$. In particular, $S$ is weakly correlated with $T_{\rm d}>90\,\rm K$ (dashed vertical line) in M17-N, and could not explain the drop of $p$ for such high $T_{\rm d}$ as shown in Figure \ref{fig:M17}.}
    \label{fig:SvsTd}
\end{figure}

Figure \ref{fig:SvsTd} shows the variation of the dispersion angle $S$ with $T_{\rm d}$ in 30 Dor (panel (a)) and in M17 (panel (b)). For 30 Dor, $S$ is poorly correlated with $T_{\rm d}$, indicating that the field tangling could not be the main explanation for the anti-correlation of $p$ vs. $T_{\rm d}$ as shown in Figure \ref{fig:30Dor_pTd}. In the case of M17, $S$ slowly varies for $T_{\rm d}>90\,\rm K$. Similarly, the variation of the magnetic field ambiguously results in the $p-T_{\rm d}$ relation in Figure \ref{fig:M17}). Via these two examples, one can see that the disruption of large grains in the vicinity of an intense radiation source is the major responsible for the decrease in $p$, instead of owing to the loss of alignment efficiency by gas collision and magnetic field tangling.

\section{Discussion}\label{sec:discussion}
\subsection{Dust polarization as a reliable tracer of magnetic fields in star-forming regions}
Magnetic fields are thought to play an important role in the evolution of the ISM and star formation. However, measurement of interstellar magnetic fields is rather difficult. One of the leading applications of dust polarization is tracing magnetic fields. Thus, the question of to what extent dust polarization can reliably trace the magnetic field is crucially important. 

As reviewed in Section \ref{sec:RAT_paradigm}, grain alignment in the ISM and MCs is dominated by the RAT mechanism. The (super)paramagnetic material (e.g., silicate grains) is magnetized through the Barnett effect, which causes the grain to have rapid Larmor precession around the magnetic field. Moreover, Barnett and inelastic relaxation effects induce efficient internal alignment of grain shortest axis with the angular momentum. The external alignment of the grain angular momentum along the ambient magnetic fields is dominated by $B-$RAT alignment because the Larmor precession is much faster than the gas randomization as well as the radiative precession. Thus, grains are aligned with their longest axes perpendicular to the magnetic field. The polarization vector of thermal dust emission is perpendicular to, whereas the polarization of starlight by dust extinction is parallel to the local magnetic field. As a result, dust polarization is a reliable tracer of magnetic fields in the diffuse ISM to molecular clouds to star-forming regions. 

Dust polarization has been used widely to measure magnetic fields in star-forming regions when combined with DCF method. This enables us to quantify the role of magnetic fields in the star formation process (see \citealt{2019FrASS...6...15P} for a recent review). Here, we discuss our recent study using dust polarization to measure magnetic fields in two star-forming regions discussed in Section \ref{sec:obs} with SOFIA/HAWC+.

\subsubsection*{{\bf (a) Mean magnetic field strength in M17}}
Knowing the variation of the magnetic fields (i.e., characterized by the polarization angle $\theta$) induced by magnetically aligned dust grains, we can estimate the field strength using the DCF method as
\begin{equation}\label{eq:DCF}
    B_{\rm mean}= f_{\rm DCF} \sqrt{4\pi \rho}\frac{\delta_{v}}{\delta_{\theta}}
\end{equation}
where $f_{\rm DCF}$ is a factor adjusting for line-of-sight and beam-integration effects, $\rho$ is the gas mass density, $\delta_{v}$ is the dispersion in velocity of non-thermal gas, and $\delta_{\theta}$ is the one for polarization angle. The dispersion of non-thermal gas is derived from spectral lines of gas tracer, while we followed an modification made by \cite{2009ApJ...696..567H} and \cite{2009ApJ...706.1504H} that one replaces $\delta_{\theta} \rightarrow D^{1/2}_{\theta}$ with $D_{\theta}$ the structure function of $\theta$ to estimate the polarization angle dispersion. We constrained the field strengths are around $980\pm 230\,\mu$G in M17-N and $1665\pm 885\,\mu$G in M17-S. The mean value of the Alfv\'enic Mach number ($\mathcal{M}_{A}$) and the mass-to-flux ratio ($\lambda$) are consequently lower than 1, indicating that M17 is sub-Alfv\'enic and sub-critical or the magnetic fields are stronger than the turbulence and able to resist against gravitational collapse. Our constraints are consistent with the low star-formation efficiency in M17 (see \citealt{2022ApJ...929...27H} for details computations).

\subsubsection*{{\bf (b) Variation of magnetic field strength in 30 Dor}}
\begin{figure}
    \centering
    \includegraphics[width=1.0\textwidth]{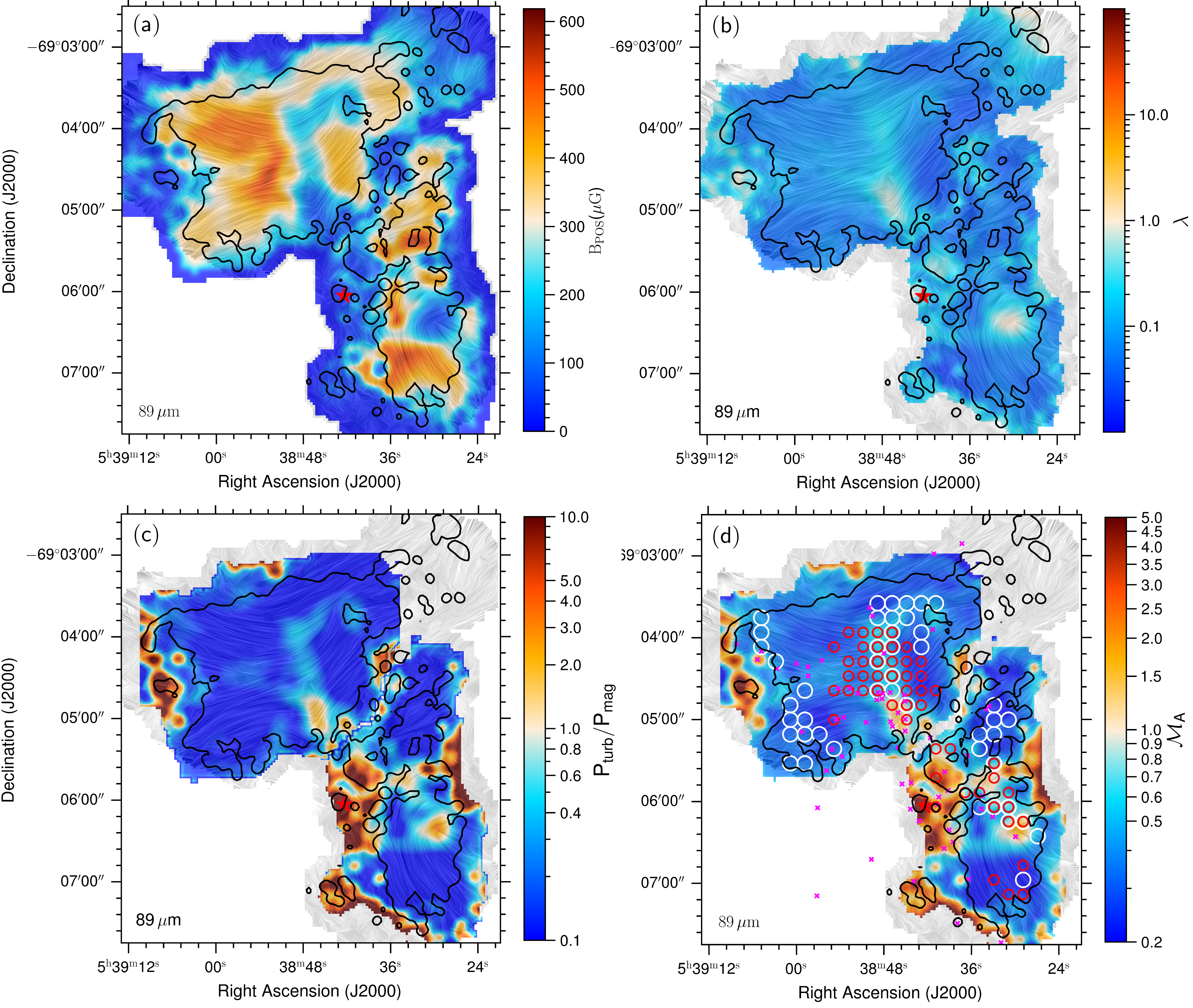}
    \caption{Maps of magnetic strength (panel a), of mass-to-flux ratio (panel b), of turbulence-to-magnetic pressure (panel c), and of Alfv\'enic Mach number (panel d) in 30 Dor (Figure adopted from \citealt{2022arXiv220512084T}). Stream lines are the magnetic field morphology inferred from thermal dust polarization. Star symbol locates the massive stellar cluster R$\,$136. Black contours enclose the area where we trust the data. Circles indicate where the gas velocity gradient parallels to the field lines (red for \textsc{CII}, and white for CO), whereas gas velocity gradient perpendiculars to the field lines. Magenta crosses are the YSOs candidates.}
    \label{fig:30Dor_Bfields}
\end{figure}

The DCF method expressed in Equation \ref{eq:DCF} can be used to make maps of magnetic field strength (see \citealt{2021ApJ...908...98G}). The maps of $\rho$, $\delta_{v}$ and $D_{v}$ are naturally required. The map of $\rho$ is due to the width of cloud, which could be determined by the half-width at half-maximum of the polarized flux autocorrelation function (see e.g., Figure 1 in \citealt{2009ApJ...706.1504H}). The map of $\delta_{v}$ is done by using a position-position-velocity datacube, while that of $D_{\theta}$ is acquired within a circular kernel of radius $w$ (see Section 3.2 in \citealt{2021ApJ...908...98G} for a critical selection of a best kernel size, $w_{\rm opt}$).

Figure \ref{fig:30Dor_Bfields} shows the case of 30 Dor as an example. The magnetic field morphology (stream lines) is inferred from SOFIA/HAWC+ observations (\citealt{2021ApJ...923..130T,2022arXiv220512084T}), which illustrates a complex but ordered structure. We constructed a map of the magnetic strength using the DCF method (panel (a)), and found a quite strong field with a maximum of few hundred micro-Gauss. Consequently, most of the material associated with 30 Dor is (1) sub-critical (i.e., $\lambda<1$, panel (b)), meaning that the magnetic field is stronger than gravity; and (2) sub-Alfv\'enic (i.e, $\mathcal{M}_{A}<1$, panel (d)), meaning that magnetic field is stronger than turbulence, which is illustrated by the turbulence-to-magnetic pressure (panel (c)). Our estimation shows that magnetic field is strong enough to support the cloud against the feedback from the intense ionization source R$\,$136. We compared the field morphology inferred from thermal dust polarization with that inferred from the velocity gradient technique (or VGT, see \citealt[and references therein]{2019NatAs...3..776H}), in which the velocity gradient is perpendicular to the ambient magnetic field. These two different techniques deliver different results at certain locations (located by the red and white circles in panel (d)) in 30 Dor. This disagreement implies that the velocity gradient is parallel to the field lines (or gas moves along the field) and gives evidence for local gravitational collapse (\citealt{2021ApJ...912....2H}). Because gas moves along the field, the gas motion is not affected by the magnetic pressure (which preferentially acts perpendicular to the field lines), and thus able to accumulate mass to trigger new stars to form in such a strong magnetic field. The locations of Young Stellar Objects are indicated by the magenta crosses for comparison.

\subsection{Can thermal dust polarization trace magnetic fields in protostellar/protoplanetary disks?}\label{sec:align_dense_gas}
In very dense regions such as protostellar cores and disks, the process of grain alignment is far more complicated than in molecular clouds, as studied in detail in \cite{Hoang.2022}. A detailed modelling of grain alignment in a protoplanetary disk by \cite{2017ApJ...839...56T} shows that very large grains tend to align along the radiation direction (k-RAT) instead of the magnetic field (B-RAT). Their study, however, evaluated the radiative precession rate for grains at low$-J$ attractors, which overestimates the efficiency of k-RAT. They assumed that all grains have {\it right} internal alignment. This assumption is likely too simplistic knowing that grain alignment can be very complicated due to a complexity of gas density. Therefore, understanding both internal and external alignments of grains becomes the key to interpret dust polarization in protostellar environments.

(Sub)millimeter polarimetry oberservations of protostars (e.g., \citealt{2014ApJ...792..116Z,2018ApJ...855...92C,Sadavoy.2019}) showed that the pattern of polarized thermal dust is complicated and diverse. Thus, we need to establish a robust understanding of grain alignment in dense regions in order to interpret observations in a reliable way (whether or not the observed patterns actually trace magnetic fields). Detailed theoretical studies (\citealt{Hoang.2022}) and numerical modeling (Giang et al. in preparation) indicate that thermal dust polarization does not always trace magnetic fields under such conditions. A diversity of polarization patterns can be the result of multiple alignment mechanisms, and inferring the magnetic field from rotating polarization vectors might lead to inaccurate conclusions (\citealt{Hoang.2022}).

\subsection{Polarization hole within the RAT paradigm}\label{sec:polhole}
In this section, we review the polarization hole effect that is frequently reported in starless cores and protostars (e.g.,\citealt{2014A&A...569L...1A}; \citealt{2015AJ....149...31J}; \citealt{2014ApJS..213...13H}; \citealt{2018ApJ...855...92C}; \citealt{2019FrASS...6...15P}) in the context of the RAT paradigm. As shown in Section \ref{sec:modelling}, the alignment size $a_{\rm min,align}$ and maximum size $a_{\rm max}$ are keys in determining the polarization degree of thermal dust. These sizes are, in turn, determined by the properties of radiation field and of environment (gas temperature and gas density). In the following we consider sources with different physical conditions, including starless cores and protostellar cores. 

For simplicity's sake, we also assume all grains have efficient internal alignment. Note that grain alignment under such conditions is much more complicated due to the complex gas density profile and interplay between different alignment mechanisms as discussed in Section \ref{sec:align_dense_gas}.

\begin{figure}
    \centering
    \includegraphics[width=0.9\textwidth]{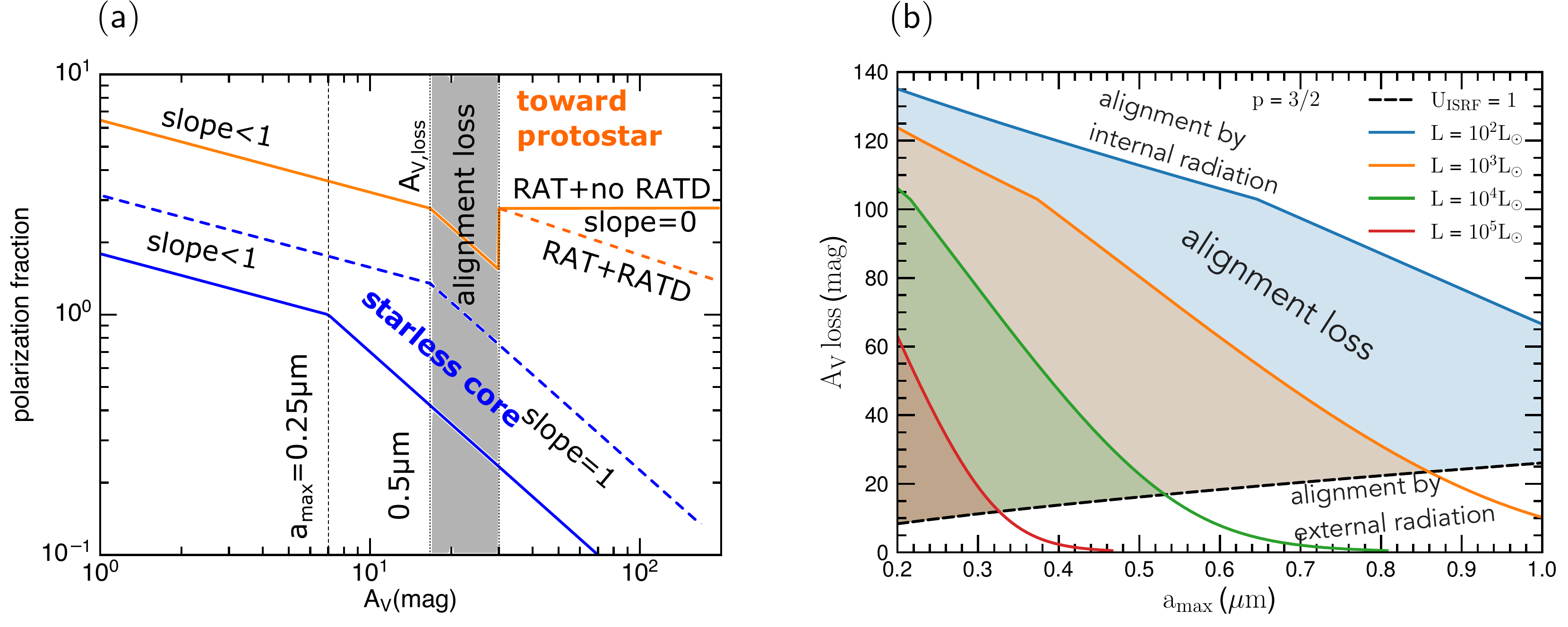}
    \caption{\textbf{(a)}: Variation of the polarization fraction (arbitrary unit) with the visual extinction expected from the RAT paradigm, described by $p\propto A^{-\xi}_{V}$. The blue lines are for a starless core with two values of the maximum grain size (solid and dashed lines). The slope changes from $\xi<1$  to $\xi=1$ when grain alignment is completely lost, highlighted by two dotted vertical lines. The orange line is for a protostellar core. Similarly, the shaded region indicates the range of $A_{V}$ where grain alignment is completely lost. Nevertheless, the internal radiation increases the alignment of grains near the source, changing the slope from $\xi=1$ to $\xi=0$ in the absence of RAT-D and to $\xi>0$ in the presence of RAT-D. Uniform magnetic field geometry is assumed. \textbf{(b)}: Variation of the $A_{V,\rm loss}$ above which grain alignment by RATs is completely lost (i.e., $a_{\rm align} \geq a_{\rm max}$), as a function of $a_{\rm max}$ for high-mass protostars. Solid lines show the results for grain alignment induced by central protostars of different luminosity. Black dashed line shows grain alignment induced by attenuated ISRF. Shaded region indicates the parameter space for complete alignment loss. Stellar temperature $T_{\ast} = 10^{4}\,\rm K$ and gas density $\propto r^{-3/2}$ are adopted. Figure adopted from \cite{2021ApJ...908..218H}.}
    \label{fig:proto_prediction}
\end{figure}

Figure \ref{fig:proto_prediction}(a) shows the expected polarization degree versus the visual extinction from the cloud surface of a starless core (blue lines) and a protostellar core (orange line) within the RAT paradigm.
\begin{itemize}
\item For a starless core (a cloud is illuminated by the interstellar radiation field (ISRF) and has no embedded sources), grain alignment is controlled by this external field. The grain alignment becomes less efficient deeper inside the cloud due to the attenuation of the ISRF and increase of gas damping. The polarization degree gradually decreases with $A_{V}$. Once the minimum aligned grain size is larger than the maximum grain size ($a_{\rm align}>a_{\rm max}$), all grains completely lost their alignment, resulting at $p\sim A^{-1}_{V}$. Thus, the location ($A_{V,\rm loss}$) where grain alignment is lost depends on $a_{\rm max}$ (vertical lines).

\item For a protostellar core, there is an internal radiation source located at within of the cloud, in addition to the external interstellar radiation field. The polarization degree similarly gradually decreases as $p\sim A^{-\rm slope}_{V}$ with a slope $<1$ before the slope becomes $1$ due to the loss of grain alignment by the ISRF. In the vicinity of the protostar, owing to the effect of the protostellar radiation, more small grains become aligned by RATs. The polarization degree is hence expected to increase monotonically (or at least stay constant when all grains are brought to align) toward the protostar. If the RAT-D mechanism occurs near the central star, the polarization degree is predicted to decrease instead of increase (dashed orange line).
\end{itemize}

Figure \ref{fig:proto_prediction}(b) shows the region from the cloud surface where grains are completely lost their alignment ($A_{V,\rm loss}$) in the case of a massive protostar. It is determined by both the internal and external radiation fields. For a given maximum grain size, the more luminous source, the narrower $A_{v,\rm loss}$, because this intense radiation field could align grains loser to the cloud surface by RATs. For a given luminosity, one can expect to have a similar narrower in range of $A_{V,\rm loss}$ for larger grain size, because the RAT alignment efficiency is proportional to grain size as shown in Section \ref{sec:St}. For a starless core, as mentioned, $A_{V,\rm loss}$ is only dependent on the maximum grain size which implies that the observed polarization hole could help to constrain grain growth in such a condition.

\subsection{Effects of RAT-A and RAT-D on grain growth and evolution, and observations}
Grain growth is thought to occur in molecular clouds where grain alignment is ubiquitous as revealed by dust polarization. Therefore, the process of grain growth must involve the growth of aligned grains. \cite{2022ApJ...928..102H} suggested that the alignment of dust grains with ambient magnetic fields has an important effect on grain growth in molecular clouds. Due to grain alignment, the process of grain growth via gas accretion and grain-grain collisions is anisotropic instead of isotropic as induced by Brownian random motions. As a result, large grains grown by gas accretion are expected to have elongation increasing with it radius. Moreover, dust aggregates formed by grain-grain collisions contain binary structures with aligned axes (\citealt{2022ApJ...928..102H}), which is recently reported in primitive interplanetary dust grains \citep{Hu.2021}.

On the other hand, the RAT-D induces the change of the grain size distribution, depending on the local radiation field and gas densities \citep{Hoang.2021a}. This disruption affects both the upper cutoff size and the abundance of smaller grains (including nanoparticles), and hence imprints on dust extinction, emission and polarization. The RAT-D effect is also important for grain surface chemistry, including rapid desorption of ice mantles, enhancement of thermal desorption of water and COMs. Therefore, the effect of RAT-A and RAT-D on the grain evolution can be accessible from observations in a large window from optical-UV to infrared to radio wavelengths (see \citealt{2020Galax...8...52H} for a review).

\subsection{Effects of RAT-A and RAT-D in the era of time-domain astrophysics}
Dust grains in the local environment of transient sources (as supernovae, kilonovae, and gamma-ray bursts) can be transiently illuminated by an intense radiation flash from the source and are transiently aligned by RAT with both high$-J$ and low$-J$ attractor (fast RAT alignment). The efficiency of fast RAT alignment and disruption is determined by $f_{\rm high-J}^{\rm fast}$. The exact value of $f_{\rm high-J}^{\rm fast}$ depends on the grain properties (shape and size) and magnetic susceptibility. For grains consisting of ordinary paramagnetic material (e.g., silicate), \cite{2021ApJ...913...63H} found that $f_{\rm high-J}$ can be about $10-70\%$ based on calculations of RATs for an ensemble of Gaussian random shapes.

Therefore, a fraction $f_{\rm high-J}^{\rm fast}$ of grains with size $a>a_{\rm disr}$ can be disrupted into fragments. 
However, the fraction $1-f_{\rm high-J}$ grains aligned at low$-J$ attractors are not disrupted by RAT-D. Therefore, in the presence of RAT-D, large grains at $a>a_{\rm disr}$ now can only rotate thermally at low$-J$ attractors. This selective disruption has an important implication for grain alignment and dust polarization. This results in a significant change in the grain size distribution and dust properties in the local environment of transients \citep{2020ApJ...888...93G,2020ApJ...895...16H}. Since the RAT-D process is time-dependent, the optical and polarization properties of transient sources would change with time just because of the change in foreground dust. This is an important effect that needs to be accounted for to infer the intrinsic properties of SNe before their luminosity and colors can be used for cosmological studies of dark energy \citep{2020ApJ...888...93G}. 

We note that the RAT-A and RAT-D effects also work for dust around other transient sources, including nova, episodic protostars, and AGN, which will be quantified in our future works.

\subsection{Mechanical Torque (MET) Alignment}\label{sec:MET}
\cite{2007ApJ...669L..77L} and \cite{2018ApJ...852..129H} realized that a gas flow could act to spin-up and align an irregular grain via the so-called Mechanical Torques (or METs). Instead of the effect of photons, as in the RAT paradigm, METs arises from the scattering of gas atoms by the grain surface that drifts through the ambient gas with the speed $v_{\rm d}$.
The suprathermal number of a grain spun-up by METs is given by (\citealt{Hoang.2022}) 
\begin{equation}
St_{\rm MET}\simeq 0.86\hat{\rho}^{1/2}s_{d,-1}^{2}a_{-5}^{3/2}\left(\frac{s^{3/2}Q_{\rm spinup,-3}}{\Gamma_{\|}}\right),\label{eq:S_MET}
\end{equation}
where $s_{\rm d}=v_{\rm d}/v_{\rm T}$ with $v_{T}$ the thermal gas velocity ($s_{d,-1}=s_{\rm d}/0.1$), and $Q_{\rm spinup,-3}=Q_{\rm spinup}/10^{-3}$ is the spin-up efficiency. Here $Q_{\rm spinup}=0$ for a spherical grain, while $Q_{\rm spinup}=10^{-6}-10^{-3}$ for irregular shapes \citep{2018ApJ...852..129H}. 

Interestingly, one can realize that $St_{\rm MET}$ is independent to the gas density, and only depends on the grain properties and drift velocity. It differs from the corresponding value for the RATs, which is less efficient in dense gas due to the gas damping. Comparing $St_{\rm MET}$ to $St_{\rm RAT}$ from Section \ref{sec:St}, we see that METs are more important than RATs for only the very small grains (e.g., $a\leq 0.02\,\mu$m for $U=10^{6}$ and $s_{\rm d}=0.5$) for low density ($n_{\rm H}=10^{3}\,\rm cm^{-3}$). In higher density ($n_{\rm H}=10^{8}\,\rm cm^{-3}$), The effect of RATs is reduced significantly so that METs become dominant mechanism to spin-up grains. Therefore, METs play an important role in dense regions (e.g., protostellar cores or disks) as shown in \cite{Hoang.2022}. Similar to RATs, grains at low$-J$ attractors induced by METs are likely to precess around the gas flow $\boldsymbol{v}$, the so-called v-MET. Whereas grains at high$-J$ attractors are likely experienced in precessing around $\boldsymbol{B}$, the so-called B-MET. Therefore, the grain alignment and induced polarization due to METs needs to be considered in modelling or simulation for a better interpretation of observations, as well as understanding physical properties of dust grain in such conditions.

\section*{Conflict of Interest Statement} 
This review uses the results from the both published and unpublished works with permissions, and the authors declare no conflict of interest.

\section*{Author Contributions}
T.H led the theory part and L.N.T led the observational data analysis part of the manuscript. L.N.T wrote the first draft of the manuscript. T.H. outlined and revised significantly the manuscript. Both authors contributed to manuscript writing, revision, read, and approved the manuscript.

\section*{Funding}
T.H. is funded by National Research Foundation of Korea (NRF) grant funded by the Korea government (MSIT) (No. 2019R1A2C1087045)

\section*{Acknowledgments}
We are grateful to the referees for helpful comments that improved our manuscript. We thank Karl Menten and Helmut Wiesemeyer for careful proof reading and comments, Nguyen Chau Giang and Nguyen Thi Phuong for useful suggestions, and Nguyen Bich Ngoc for allowing us to use her preliminary results.

\bibliographystyle{Frontiers-Harvard} 

\end{document}